\documentclass[aps,pra,twocolumn,superscriptaddress,amssymb,amsmath,floatfix,longbibliography,english,nofootinbib]{revtex4-2} %
\usepackage[T1]{fontenc}
\usepackage{upgreek}
\usepackage{pgfplots}
\usepackage{alphalph}
\usepgfplotslibrary{groupplots}
\usetikzlibrary{plotmarks}
\usepackage{mathtools} %
\usepackage[mathscr]{euscript}
\usepackage{natbib}
\usepackage{braket}
\usepackage{graphicx}
\usepackage[bookmarks=false]{hyperref}
\usepackage{xcolor, soul}
\sethlcolor{yellow}
\usepackage{booktabs} %
\usepackage{multirow}
\usepackage{array} %
\usepackage{makecell} %
\usepackage[nameinlink]{cleveref}
\definecolor{myred}{HTML}{EE351E}
\definecolor{mypur}{HTML}{995CB3}
\definecolor{mybla}{HTML}{211E1E}
\definecolor{myblu}{HTML}{234AA3}
\definecolor{myora}{HTML}{F28C28}
\definecolor{prxblue}{HTML}{2E358F}
\definecolor{mybeige}{HTML}{00C2D1}
\definecolor{mygreygreen}{HTML}{598B2C}
\definecolor{mygreyblue}{HTML}{596475}
\makeatletter
\pgfplotsset{
    compat=1.5,
    tick label style={font=\normalsize},
    label style={font=\normalsize},
    legend style={font=\footnotesize},
    width=8.73cm,
    height=6.23cm,
    xtick pos=left,
    ytick pos=left,
    auto title/.style={title=(\alphalph{\pgfplots@group@current@plot})},
    every axis title/.append style={at={(0.08,0.80)}, font=\large},
    legend style={rounded corners},
}
\makeatother

\colorlet{mycolor}{myblu}
\hypersetup{
    colorlinks = true,
    citecolor = mycolor,
    linkcolor = mycolor,
    urlcolor =  mycolor,
}

\newcommand{\synd}[4]{|\!#1\!#2\!#3\!#4\rangle}

\newcommand{\figref}[2]{\cref{#1}\textcolor{mycolor}{#2}}
\newcommand{\methodsref}[1]{Methods \ref{#1}}
\def\supplmat#{\href{https://www.overleaf.com/project/635be6d9167e6a2caa862085}{supplemental material}}
\def\supplmatlink#{\href{https://www.overleaf.com/project/635be6d9167e6a2caa862085}{overleaf.com/project/635be6d9167e6a2caa862085}}
\newcommand{\CNOT}{\textsc{CNOT}}
\newcommand{\CPHASE}{\textsc{CPHASE}}
\newcommand{\epsilontrans}{\epsilon_\mathrm{trans}}

\newcommand{\epsiloneff}{\epsilon_\mathrm{eff}}
\newcommand{\epsilonr}{\epsilon_\mathrm{r}}
\newcommand{\ptrans}{p_{\mathrm{trans}}}
\newcommand{\mI}{m_\mathrm{I}}
\newcommand{\mII}{m_\mathrm{II}}
\newcommand{\etae}{\eta_\mathrm{e}}
\newcommand{\etad}{\eta_\mathrm{d}}
\newcommand{\typei}{\textsc{type~i}}
\newcommand{\typeii}{\textsc{type~ii}}
\newcommand{\Typei}{\textsc{Type~i}}

\newcommand{\SKR}{S\!K\!R}

\newcommand{\mytextllbracket}{[\kern-0.16em[}
\newcommand{\mytextrrbracket}{]\kern-0.16em]}

\newcommand{\textnkd}[1]{\mytextllbracket{#1}\mytextrrbracket}
\newcommand*\redspin{\vcenter{\hbox{\includegraphics[width=0.78em]{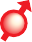}}}\mkern-1.6mu}
\newcommand*\purplespin{\vcenter{\hbox{\includegraphics[width=0.78em]{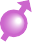}}}\mkern-1.6mu}
\newcommand*\greenspin{\vcenter{\hbox{\includegraphics[width=0.78em]{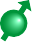}}}\mkern-1.6mu}

\begin{document}

\clearpage{}%
\author{Kah Jen Wo}
\email{kjwo@outlook.sg}
\affiliation{QuTech, Delft University of Technology, Lorentzweg 1, 2628 CJ Delft, The Netherlands}
\affiliation{Quantum Technologies, National University of Singapore, Queenstown 117543, Singapore}

\author{Guus Avis}
\email{guusavis@hotmail.com}
\affiliation{QuTech, Delft University of Technology, Lorentzweg 1, 2628 CJ Delft, The Netherlands}
\affiliation{Quantum Computer Science, EEMCS, Delft University of Technology, Lorentzweg 1, 2628 CJ Delft, The Netherlands}
\affiliation{Kavli Institute of Nanoscience, Delft University of Technology, Lorentzweg 1, 2628 CJ Delft, The Netherlands}

\author{Filip Rozp{\k{e}}dek} %
\affiliation{Pritzker School of Molecular Engineering, University of Chicago, Chicago, IL 60637, USA}
\affiliation{College of Information and Computer Sciences, University of Massachusetts Amherst, Amherst, Massachusetts 01003, USA}

\author{Maria Flors Mor-Ruiz}
\affiliation{Universit{\"a}t Innsbruck, Institut f{\"u}r Theoretische Physik, Technikerstra{\ss}e 21a, 6020 Innsbruck, Austria}

\author{Gregor Pieplow}
\affiliation{Department of Physics, Humboldt-Universit\"{a}t zu Berlin, Newtonstra{\ss}e 15, 12489 Berlin, Germany}

\author{Tim Schr\"{o}der}
\affiliation{Department of Physics, Humboldt-Universit\"{a}t zu Berlin, Newtonstra{\ss}e 15, 12489 Berlin, Germany}

\author{Liang Jiang} %
\affiliation{Pritzker School of Molecular Engineering, University of Chicago, Chicago, IL 60637, USA}

\author{Anders S.\ S{\o}rensen} %
\affiliation{Center for Hybrid Quantum Networks (Hy-Q), The Niels Bohr Institute, University of Copenhagen, Blegdamsvej 17, DK-2100 Copenhagen {\O}, Denmark}

\author{Johannes Borregaard}
\email{j.borregaard@tudelft.nl}
\affiliation{QuTech, Delft University of Technology, Lorentzweg 1, 2628 CJ Delft, The Netherlands}
\clearpage{}%

\title{Resource-efficient fault-tolerant one-way quantum repeater with code concatenation}

\date{October 7, 2023}

\clearpage{}%
\begin{abstract}
    One-way quantum repeaters where loss and operational errors are counteracted by quantum error correcting codes can ensure fast and reliable qubit transmission in  quantum networks.
    It is crucial that the resource requirements of such repeaters, for example, the number of qubits per repeater node and the complexity of the quantum error correcting operations are kept to a minimum to allow for near-future implementations.
    To this end, we propose a one-way quantum repeater that targets both the loss and operational error rates in a communication channel in a resource-efficient manner using code concatenation.
    Specifically, we consider a tree-cluster code as an inner loss-tolerant code concatenated with an outer 5-qubit code for protection against Pauli errors.
    Adopting flag-based stabilizer measurements, we show that intercontinental distances of up to 10,000 km can be bridged with a minimized resource overhead by interspersing repeater nodes that each specializes in suppressing either loss or operational errors.
    Our work demonstrates how tailored error-correcting codes can significantly lower the experimental requirements for long-distance quantum communication.
\end{abstract}
\clearpage{}%

\maketitle

\section{Introduction}

The ability to faithfully transmit quantum information over long distances opens up new opportunities for secure communication~\cite{Lo2014,Xu2020}, sensing networks~\cite{Zhang2021} and distributed quantum computing~\cite{Buhrman2003}. There has been impressive progress towards the realization of quantum networks with the demonstration of long-distance entanglement distribution through satellites~\cite{Yin2017}, memory-enhanced quantum communication~\cite{Bhaskar2020}, and a multi-node quantum network~\cite{Pompili2021}. Quantum communication over intercontinental distances, however, remains a formidable challenge due to attenuation and degradation of the quantum signal as a result of loss and operational errors.

There are, in general, two classes of quantum repeater architectures that have been proposed to overcome these obstacles~\cite{Munro2015,Muralidharan2016}. Two-way repeaters divide the total distances into smaller links where heralded entanglement can be created in a probabilistic manner by direct photon transmission and success is heralded by two-way communication between the repeater nodes~\cite{Briegel1998,Duan2001,Sangouard2011}. Such architectures require long-lived multi-mode quantum memories to reach high communication rates~\cite{Simon2007}. An alternative approach, which is the focus of this work, is to encode the quantum information in quantum error-correcting codes to battle both loss and operational (Pauli) errors as the signal is being transmitted from one repeater node to the next in a one-way architecture~\cite{Munro2012,Muralidharan2014,Ewert2016,Borregaard2020,Rozpedek2021,Fukui2021}. Such fault-tolerant one-way repeaters allow bridging arbitrary distances with high communication rates since the repetition rate is set by the local processing time of the repeater nodes. However, they often have daunting requirements in terms of resources needed for the physical implementation of the repeater \cite{Muralidharan2014,Ewert2016,Lee2019,Ewert2017}.
Focusing on the dominant error of photon loss and abandoning fault tolerance for operational errors can significantly relax these requirements as recently demonstrated in Ref.~\cite{Borregaard2020}.
In particular, it was shown that by using only loss-tolerant photonic tree-cluster encoding, the bottleneck of transmission loss could effectively be overcome with only three spin qubits per repeater node.
That approach is, however, not fault-tolerant against operational errors, and as a result the operational error experienced by the qubit in transmission must be very low ($\sim \!10^{-5}$) in order to bridge intercontinental distances.

Alternatively, the concatenation of two different quantum error correction codes has been considered to ensure fault-tolerant operation and efficiently address both loss and operational errors in quantum repeaters. In particular, the concatenation of a continuous-variable (CV) Gottesman-Kitaev-Preskill (GKP) code with small discrete-variable (DV) codes has recently been proposed~\cite{Rozpedek2021,Schmidt2022, Fukui2021, rozpkedek2023all, schmidt2023error}, and was shown to have increased communication performance while reducing the resource cost.
While the experimental generation of optical GKP states has recently been demonstrated \cite{Konno2023}, the construction of such states with sufficiently high quality remains a daunting challenge due to the requirements for the efficiency and performance of the hardware.

\begin{figure*}[t]
    \centering
    \includegraphics[width=0.95\linewidth]{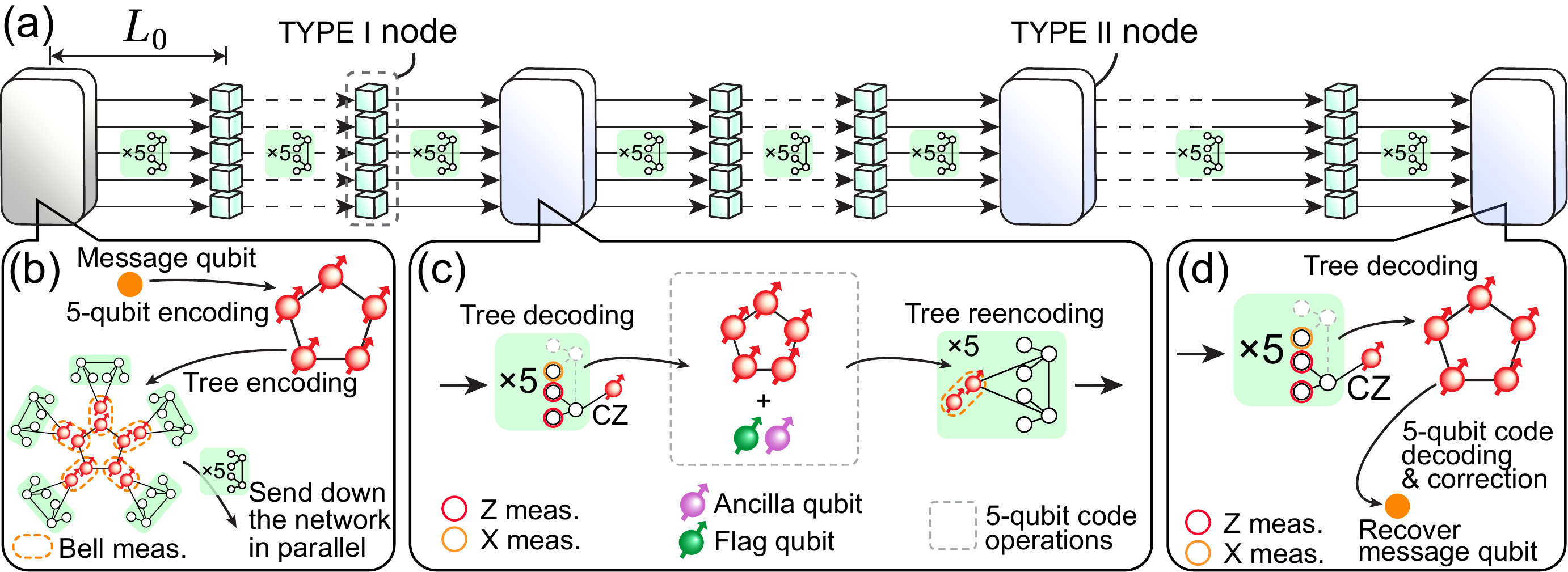}
    \caption{
        (a) Overview of the hybrid one-way quantum repeater network containing two types of repeater nodes: \typei{} and \typeii{}.
        (b) In the start node (greyed because no error correction is involved), a message qubit is encoded using the 5-qubit~code into five data spin qubits.
        Then, each of those data spin qubits is encoded in parallel into a photonic tree-cluster state via a Bell state measurement (orange dashed box) with the root spin qubit of the tree. The $[2,2]$ tree is  used as an example here.
        The trees (each encased in a green box) are then sent in parallel along the repeater network where the nodes a distance $L_0$ apart from each other.
        \textsc{Type i} nodes consist of five processing blocks. Each of these blocks decodes the received tree and re-encodes the decoded qubit into a fresh tree via heralded storage \cite{Borregaard2020}.
        (c) \textsc{Type ii} nodes decode the incoming tree at the tree level, then perform stabilizer operations (i.e., two-qubit gates and syndrome extraction on the 5-qubit~code level) between the decoded qubit and the ancilla qubits with accompanying flag qubits in the nodes.
        After the stabilizer operations, the decoded qubit is re-encoded back into a new tree and sent off to the next node.
        (d) At the end node, the incoming five trees are received and decoded in parallel.
        The 5-qubit~code corrections are applied according to the syndromes obtained along the network and finally they are decoded back into the message qubit.
    }
    \label{fig:network-overview}
\end{figure*}

In this article, we propose a purely DV-based one-way quantum-repeater architecture (see \cref{fig:network-overview}) that uses code concatenation and flag-based quantum error correction to achieve fault-tolerant operation in a resource-efficient manner.
Specifically, we combine a loss-tolerant tree cluster state as an inner code \cite{Borregaard2020} with a 5-qubit code operated with a flag qubit as an outer code.
The generation of large photonic cluster states can be done efficiently with single emitters~\cite{Buterakos2017} and has been demonstrated with a number of experimental platforms~\cite{Thomas2022,Cogan2023,Coste2023,Appel2022}.
The 5-qubit code \cite{Laflamme1996} ensures efficient correction of operational errors for which the tree code does not provide protection. By adopting flag-based stabilizer measurements, we show that communication rates in the kHz range can be achieved over thousands of kilometers for qubit transmission errors $\sim\!10^{-3}$ with a minimized qubit overhead per repeater node.

Our design intersperses two types of repeater nodes that operate differently on the two codes similarly to the CV-DV architecture of Ref.~\cite{Rozpedek2021}.
\Typei{} nodes perform error correction solely on the tree code while \typeii{} nodes perform error correction on both codes. Consequently, \typeii{} nodes are more complex than \typei{} nodes. We show how the extra cost of \typeii{} nodes can be included in the design of the repeater architecture to maximize network performance while minimizing the cost. Finally, we outline how both types of repeater nodes can be constructed from 5 modular processors containing only 1 quantum emitter and at most 4 memory spins each. This makes our design suitable for implementation with current quantum-network hardware such as solid-state defect centers coupled to a small qubit registers of nuclear spins \cite{Stas2022,Nguyen2019}.

The paper is structured as follows. In \cref{sec:protocol}, we introduce the high-level repeater protocol and its building blocks. In \cref{sec:experimental-considerations}, we discuss a possible implementation with efficient spin-photon interfaces based on cavity-coupled quantum emitters and in \cref{sec:performance}, we quantify the performance of the repeater, which we optimize for various distances in \cref{sec:results}.

\section{Quantum repeater protocol}\label{sec:protocol}
The repeater architecture incorporates three main steps. First, a message qubit is encoded in an error-correction code at the starting node.
This is followed by re-encoding and error correction at subsequent repeater nodes before a final decoding at the end node. A high-level schematic of the entire repeater network is shown in \cref{fig:network-overview}, and we will now address each of the three steps separately in more detail below. %

\subsection{Encoding}\label{subsec:encoding}
The message qubit at the start node is first encoded into the \textnkd{5,1,3}~code, otherwise known as the 5-qubit~code (see \cref{fig:network-overview}\textcolor{mycolor}{b}).
Here, \textnkd{\textit{n},\,\textit{k},\,\textit{d}} refers to a code that encodes $k$~logical~qubits using $n$~physical~qubits with a code distance $d$.
We have chosen the 5-qubit code as the outer code because it is the smallest quantum error correcting code that can correct single arbitrary Pauli errors \cite{Laflamme1996}, an attractive trait in minimizing physical resource requirements.
We note that the encoding of the logical state can be performed fault-tolerantly using a scheme recently demonstrated with a diamond Nitrogen-Vacancy platform in Ref.~\cite{Abobeih2022}.
This involves heralding the desired encoded logical state via repeated stabilizer measurements in conjunction with a flag qubit.

\begin{figure}[b!]
    \centering
    \includegraphics[width=0.66\linewidth]{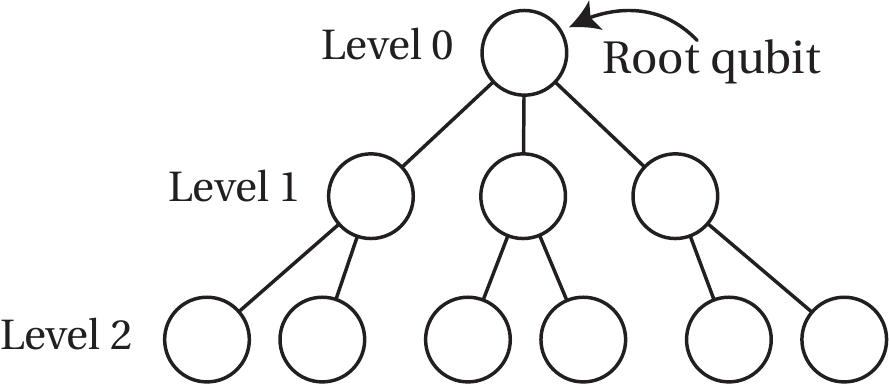}
    \caption{A $[3,2]$ tree-cluster state is shown as an example to illustrate the different levels of a tree. Each vertex/circle represents a qubit initially prepared in the state $\ket{+}=(\ket{0}+\ket{1})/\sqrt{2}$. The edges connecting the vertices correspond to a \CPHASE{} gate being applied between the two qubits.}
    \label{fig:tree}
\end{figure}

After encoding the message qubit into the outer 5-qubit code, each of the five data qubits of the code are then further encoded into a photonic tree-cluster state (inner code) that provides loss tolerance via information redundancy~\cite{Varnava2006}.
The tree-cluster code can be described by its branching vector $\vec{t}=[b_0,b_1,\ldots,b_d]$ which determines the degree of branching from each level in the tree beginning from the root qubit (see \cref{fig:tree}).
Every node in the tree cluster state described by the branching vector represents a qubit in state $|+\rangle=(\ket{0}+\ket{1})/\sqrt{2}$ and each of the edges represents \CPHASE{} gate. This resulting tree cluster state is then the unique state that is stabilized, i.e., it has eigenvalue $+1$ for each of the operators $K_\nu=X_\nu\prod_{b\in\mathscr{N}_\nu}Z_b$, where $\nu$ labels the qubits of the tree state and $\mathscr{N}_\nu$ denotes the set of qubits connected to the $\nu^\mathrm{th}$ qubit.
To encode a data qubit in the tree-cluster code, a single Bell state measurement of the root qubit of the tree-cluster state and the respective data qubit is sufficient as detailed in Ref.~\cite{Borregaard2020} and shown in  \figref{fig:network-overview}{b}.

At this point, the message qubit has been encoded into 5 photonic tree-cluster states, which are sent down the repeater network in parallel (see \cref{fig:network-overview}\textcolor{mycolor}{a}).
Note that we are assuming a generation scheme of the photonic tree-cluster states where each photon of the tree is generated sequentially from a single emitter featuring a multi-level electronic ground state manifold \cite{Buterakos2017}, hence the photons are sent down the optical fibers using time-bin encoding. For details on this scheme, we refer to Ref.~\cite{Borregaard2020}.
Consequently, the 5 trees can be transmitted using 5 single-mode fibers.

\subsection{Re-encoding and error correction}\label{subsec:re-encoding}
In the repeater network, the incoming photonic qubits from the previous node will be received by either a \typei{} or \typeii{} repeater node, each specializing in dealing with types of errors.
The \typei{} nodes only operate on the inner tree code to correct transmission loss, and we envision that each \typei{} node consists of 5 parallel repeater subnodes similar to the nodes considered in Ref.~\cite{Borregaard2020}.
Each of these 5 subnodes receives and operates on 1 of the 5 tree-encoded qubits of the outer 5-qubit code.
Their basic operation is to decode each incoming photonic tree-encoded data qubit into a spin-qubit and then re-encode it into a new photonic tree encoding.
In this process, loss will be corrected but not logical errors on the outer 5-qubit code level.

The re-encoding of the data qubit into the new tree-cluster state is achieved with a Bell state measurement between any one of the first-level photonic qubits in the incoming tree and the root qubit of the new tree along with the measurement of the remaining qubits in the incoming tree in appropriate single-qubit bases according to the stabilizer generators of the tree (see \cref{subsec:encoding}).
The whole procedure can be performed in a loss-tolerant way with only two memory spins and a single cavity-coupled emitter as outlined in Ref.~\cite{Borregaard2020}.
Note that we assume that decoherence errors in all memory spins considered in this article is negligible compared to other gate errors because the information is only stored short-term in the memory spins in our network protocol.

Importantly, the Bell state measurement between a first-level photonic qubit of the incoming tree and the stationary root qubit of the new tree is not protected against faulty gate operations, which introduces re-encoding error $\epsilonr$ on our encoded qubit on the tree code level.
In addition, we assume that each qubit in the trees is subjected to a single qubit depolarizing error of $\epsilon_0$, which comes from the inherent operational error in the underlying hardware.
The presence of this re-encoding error is why the scheme in Ref.~\cite{Borregaard2020} is not fault-tolerant and requires very low re-encoding error rates for long-distance communication.
Here, we solve this issue with the outer 5-qubit code, which provides an extra layer of protection for the message qubit in the \typeii{} repeater nodes.

The \typeii{} repeater nodes are designed to primarily correct for the accumulated re-encoding errors in the network by operating on both the inner tree encoding and the outer 5-qubit encoding.
We refer to this as the accumulated re-encoding~error between \typeii{} nodes as the transmission~error $\epsilontrans$.
This transmission error experienced by the encoded qubit is modeled via the single qubit depolarizing channel discussed in \methodsref{methods:subsec:error-model}.
The first step to correct for $\epsilontrans$ is to decode the tree encoding using the same procedure as the \typei{} nodes thereby correcting loss errors.
Once the 5 incoming trees are decoded into 5 single spin qubits, the syndrome extraction of the 5-qubit code is performed in a fault-tolerant manner using a flag qubit to correct for $\epsilontrans$. For the fault-tolerant error correction protocol, refer to \methodsref{methods:subsec:error-correction}.
After the syndrome extraction, the decoded qubits are re-encoded back into newly generated trees and are sent down the network.
Note that since a \typeii{} node also performs the same decoding and encoding procedure as a \typei{} node, we also take into account the errors generated by these steps in the transmission error. Refer to \methodsref{methods:subsec:error-model} for how these are taken into account.

Additionally, we consider noisy two-qubit gates in \typeii{} nodes, which we model via a two-qubit depolarizing channel with an error rate of $\epsilon_0$ discussed in \methodsref{methods:subsec:error-model}.
Note that, as we will discuss in \cref{sec:experimental-considerations}, there are two-qubit gates in each of the \typeii{} nodes that have to be performed in a teleported manner, which have an error rate of $3\epsilon_0$ because each of them is essentially comprised of 3 two-qubit gates.

In \cref{sec:results}, we discuss that the inherent operational error is related to the re-encoding error via the relation $\epsilonr\approx3\epsilon_0$ found through our numerical simulation (see \methodsref{methods:subsec:error-model}).
This means with the increase of the re-encoding error, the two-qubit gates in \typeii{} nodes become noisier and that placements of \typeii{} nodes in the network would become more sparse.
Note that we assume errors introduced by single qubit gates are negligible since they are typically much smaller than errors induced by two-qubit gates in an architecture based on diamond defects \cite{Rong2015}.
We also assume that the dominant errors in the qubit readout operations enter through the noisy two-qubit gates because the qubit readout operations considered in this paper are always associated with stabilizer operations of the 5-qubit~code.
Therefore, the qubit readouts themselves are considered error-free.
In summary, the faulty re-encoding procedure in both \typei{} and \typeii{} nodes, and the noisy two-qubit gates in the \typeii{} nodes are the only source of operational errors in our model.

Besides correcting for transmission errors, the 5-qubit code can also correct for failed decoding attempts of the photonic tree cluster states due to loss of too many photons. In general, any quantum error-correcting code capable of correcting $t$ arbitrary Pauli errors could also be used to correct $2t$-erasure errors~\cite{Grassl1997}.
Consequently, the 5-qubit~code can correct for up to two failed decoding attempts (erasure errors) of the incoming photonic tree-states. While the probability of 2-erasure errors occurring within the same \typeii{} node is very small due to the loss correction of the \typei{} nodes, we find that correcting for a single lost tree, i.e., 1-erasure errors, at the \typeii{} nodes can significantly boost the communication rate (see \supplmat{}). We therefore apply the erasure-error correction in this work in \cref{sec:performance}, for more details please refer to \methodsref{methods:subsubsec:erasure-error-correction}.

\subsection{Decoding}

The extracted syndromes of the error correction, which provide information about whether a 1-erasure error has occurred or not, are simply stored in a classical register. This register is classically communicated to the end node where it is interpreted and only when required single-qubit gates are applied to correct the errors according to the interpreted syndromes. Thus, it is not necessary to perform any correction on the quantum level at the repeater nodes. This allows to both avoid additional decoherence due to latency introduced by applying the error correction at every \typeii{} node and allows for performing corrections based on the joint information of all error syndromes from the repeater network. This approach is known as \textit{Pauli frame updating} \cite{Knill2005}.

The end node is a \typeii{} node, where the tree-cluster states are first decoded into the 5 data spin qubits. Error correction is then performed on the 5 data spin qubits according to the syndromes extracted throughout the network, after which they are then further decoded back into the original message qubit and measured. Note that since the encoding and decoding procedure on the 5-qubit code level is only done at the start and end node, we assume that errors introduced at these steps are negligible.

\section{Implementation}\label{sec:experimental-considerations}

We will now discuss a specific modular design of the repeater nodes that allows for an implementation with small qubit processors containing 1 cavity-coupled quantum emitter and at most 4 memory spins.
Our choice of a modular design is motivated by current hardware capabilities with e.g. solid state defect centers where efficient spin-photon interfaces can be achieved through coupling to nanophotonic resonators~\cite{Nguyen2019,Rugar2021,Stas2022}.
By coupling to a few near-by nuclear spins such small processors can be envisioned.
Coupling between the processors (needed for \typeii{} nodes) can be achieved through photon-mediated interactions between the emitters as we outline below.

As described above, a \typei{} node consists of 5 parallel repeater subnodes similar to the node described in Ref.~\cite{Borregaard2020}, with each subnode requiring only 2 memory spins and 1 quantum emitter for the generation of depth-3 photonic tree-cluster states to perform the loss-tolerant re-encoding operation.

The photonic tree generation follows the scheme of Ref.~\cite{Buterakos2017}, where the photons of the tree are sequentially emitted from the emitter with intermediate entangling operations between the emitter and the 2 memory spins. In this way the branches of the tree are generated one by one starting from the bottom. We note that this generation scheme requires a subsequent re-ordering of the photons using optical switching and a delay line as detailed in Ref.~\cite{Borregaard2020} such that the first-level photons lead when arriving at the next repeater node. This is necessary since the absence/presence of a first-level photon determines the measurement bases of the remaining photons of the branch.

A key element for the re-encoding operation is a cavity-mediated CPHASE gate between an incoming photon and the emitter~\cite{Reiserer2014,Tiecke2014,Kalb2015,Sun2018}. The basic principle of this gate is that if only one of the ground states of the emitter is coupled to the cavity mode, an incoming photon will be reflected with/without a $\pi$-phase shift if the emitter is in the uncoupled/coupled state. This operation allows for transferring the quantum state of a photon to the emitter, heralded by subsequent detection of the photon.
As shown in Ref.~\cite{Borregaard2020}, this makes the re-encoding operation robust to transmission losses by first transferring the state of a first-level photon of the incoming tree to the emitter in a heralded way followed by a Bell state measurement between the emitter and a memory spin. The memory spin constitutes the root qubit of a fresh tree cluster state emitted prior to the reception of the incoming tree.

\begin{figure}[b!]
    \centering
    \includegraphics[width=\linewidth]{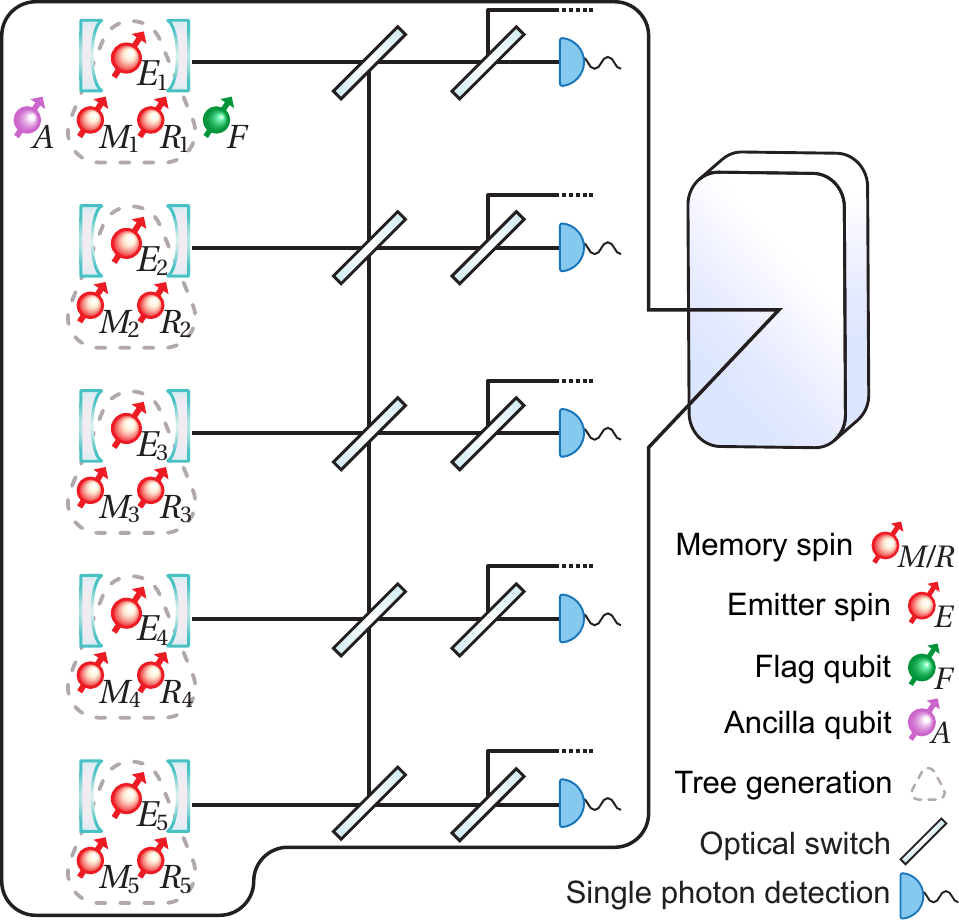}
    \caption{
        The physical resources of the \typeii{}~repeater~nodes: 5 sets of single-sided cavities, each with a quantum emitter $\redspin_{E_j}$ with $j\mkern-3mu\in\mkern-3mu\{\mkern-1mu1,\mkern-1.2mu2,\mkern-1.2mu3,\mkern-1.2mu4,\mkern-1.2mu5\mkern-1mu\}$ and accompanying memory spins.
        The 5 emitter spins host the decoded qubits from the 5 trees. They are equipped with optical switches that allow for performing teleported \CNOT{} gates between the desired decoded qubits that are physically far apart.
        In all the sets, 2 extra qubits, i.e., $\redspin_{M_j}$ and $\redspin_{R_j}$, are needed for tree generation and teleported two-qubit gates.
        In one of the sets, another 2 qubits are needed for the ancilla and flag qubits, which are shown as $\purplespin_{A}$ and $\greenspin_{F}$, respectively.
    }
    \label{fig:inside-type-ii-detailed}
\end{figure}

The requirements of \typeii{} nodes are different from the \typei{} nodes since they perform error correction on the outer 5-qubit code in addition to the loss correction on the inner tree-cluster code. Here we outline how a modular architecture can enable this increased functionality with a very modest increase in resources compared to \typei{} nodes.
Similar to \typei{} nodes, we consider an implementation with 5 cavity-coupled emitters that each can receive and generate their tree-cluster states using 2 additional memory spins per emitter (see~\cref{fig:inside-type-ii-detailed}).
To allow for the syndrome extraction of the 5-qubit code, we only need one of these emitters to be coupled to 2 extra memory spins, which act as ancilla and flag qubits of the 5-qubit code.
Consequently, we require a sequential extraction of the syndromes, which increases the duration of the error-correction procedure. A faster parallel extraction would be possible with the added expense of more ancilla and flag qubits, but here we choose to focus on the sequential operations to minimize the number of qubits per repeater node.

\begin{figure}[b!]
    \centering
    \includegraphics[width=0.97\linewidth]{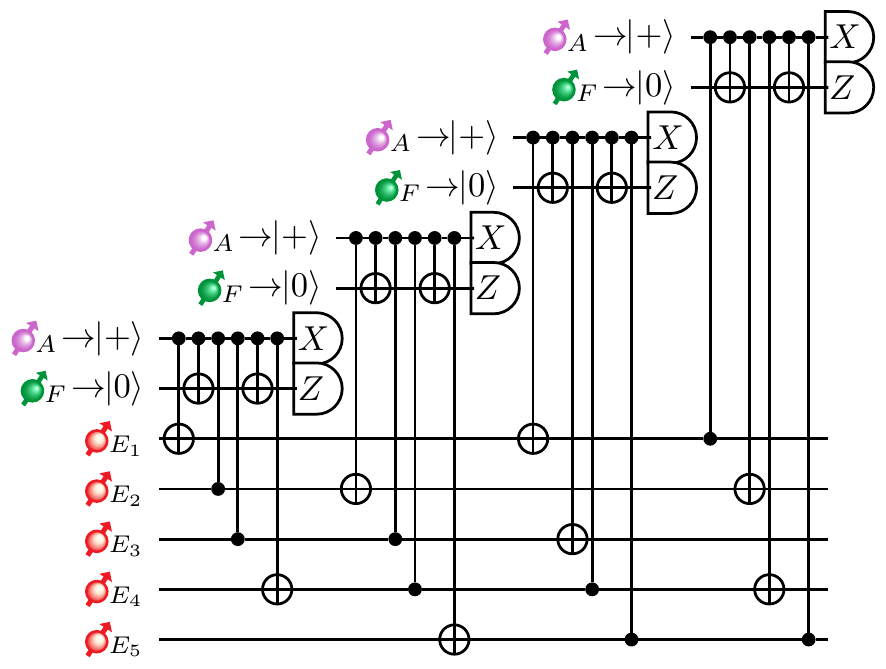}
    \caption{The circuit that a \typeii{} node performs to fault-tolerantly correct for operational errors using the 5-qubit code. For the error correction protocol, refer to \methodsref{methods:subsec:error-correction}.}
    \label{fig:flag_qcircuit}
\end{figure}

We assume direct coupling between an emitter and its nearby memory spins based, for example, on spin-spin interactions, which allows for the implementation of multi-qubit gate operations. However, as shown in \cref{fig:flag_qcircuit}, the syndrome extraction of the 5-qubit code requires two-qubit gates between the ancilla qubits belonging to different emitter-cavity systems and the data spin qubit. To implement such non-local operations, teleported gates can be used.

\begin{figure}[h]
    \centering
    \includegraphics[width=0.85\linewidth]{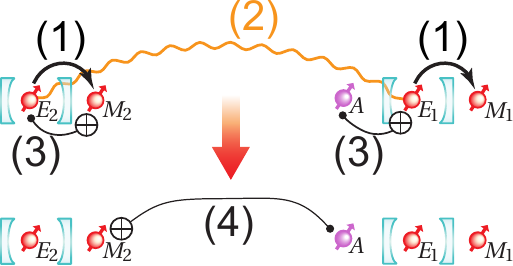}
    \caption{
        Procedure for performing a teleported \CNOT{} gate between an ancilla qubit (control) and a data qubit (target) by heralding a Bell pair between two emitters and performing a measurement on the Bell pair.
        For simplicity, not all spins coupled to each cavity are shown (\cref{fig:inside-type-ii-detailed}).
        We use the realization of a teleported \CNOT{} gate between an ancilla qubit $\purplespin_A$ and the decoded qubit in $\redspin_{E_2}$ as an example.
        The steps are as follows:
        (1)~The decoded qubits in the emitter spins $\redspin_{E_j}$ are transferred to the auxiliary memory spins $\redspin_{M_j}$.
        (2)~A Bell pair is heralded between the two emitter spins.
        (3)~Local spin-spin \CNOT{} gates are performed between spins of interest.
        (4)~The qubits corresponding to the Bell pair heralded in step (2) are measured to perform a teleported \CNOT{} gate between desired control and target qubits up to some Pauli gate(s) correction according to the measurement outcome.
    }
    \label{fig:teleported-cnot}
\end{figure}

The teleported \CNOT{}~gate involves heralding a Bell pair between two emitters via the transmission and subsequent reflection of a photon \cite{Gottesman1999,Jiang2007,Chou2018,Wan2019} (see \cref{fig:teleported-cnot}).
The creation of such a Bell pair can be performed using the same operations and same spins, which are shown enclosed in dashed gray triangles in \cref{fig:inside-type-ii-detailed}, to decode an incoming tree-cluster and generate a new tree-cluster.
Note that we can only perform the Bell pair creation once the fresh tree has been generated with the memory spin $R$ hosting the root qubit and the emitter spin is ``freed'', i.e., reinitialized.
Specifically, for creating a Bell pair, a photon entangled with one of the emitters is generated by emission and then scattered off another cavity-spin system followed by a heralding measurement of the photon.
This will prepare the two emitter spins in a Bell state, which can be used to mediate a gate between two remote spin qubits through gate teleportation.
To perform teleported~\CNOT{}~gates between two qubits of interest, local \CNOT{}~gates between the qubits of interest and the emitter spins are performed followed by single-qubit measurements of the emitter spins.
Note that the emitter spins in the repeater nodes hold the decoded data qubits upon the reception of trees, therefore we need to first free them up by transferring the decoded data qubits onto the auxiliary memory spins as shown in \cref{fig:teleported-cnot} before performing the \CNOT{} gate.
This will then amount to a \CNOT{}~operation between the two distant spin qubits up to a single qubit correction dictated by the measurement outcome.
In order to connect the pairs of cavity-emitter systems dictated by the error-syndrome-extraction circuit (see \cref{fig:flag_qcircuit}), we imagine that fast optical switching~\cite{Vedovato2018} is employed to ensure the generation of Bell pairs between the respective emitter spins.

\section{Repeater performance}\label{sec:performance}
In this section, we benchmark the performance of the repeater network by looking at the secret key rates that are achieved when executing a Quantum Key Distribution (QKD) protocol. This quantity encapsulates both the fidelity of the final message qubit and the raw bit rate of the network. Therefore, the determination of the secret key rate provides an excellent means of assessing the general performance of the network.
Since our repeater network is based on the one-way quantum repeater protocol, we consider the prepare-and-measure-based six-state protocol \cite{Bruss1998} as the most suitable QKD protocol, whose secret key fraction is dependent on the effective error rate of the message qubit at the end node. Details about this secret key fraction is given in \methodsref{methods:subsec:repeater-performance}.

Furthermore, we boost the secret key rate by leveraging the 5-qubit code's ability to correct for erasure errors as explained in \cref{subsec:re-encoding}.
Note that our secret key rate depends on 2 facts: (1) the effective error rate at the end node depends on how many erasure errors occurred in the network, and (2) the presence of erasure errors results in a different successful transmission probability of the message qubit through the network (see \methodsref{methods:subsec:repeater-performance}).
Recall that we only consider correcting for 1-erasure errors and not 2-erasure errors at the 5-qubit code level for reasons explained in \cref{subsec:re-encoding}.
These dependencies are accounted for in the secret key rate calculation by performing a weighted sum over the possible number of 1-erasure error occurrences in the network, effectively resulting in an average secret key rate
\begin{equation}\label{eqn:SKR-secret-key-rate}
    \SKR=\tau_\mathrm{tot}^{-1}\sum^{\mII}_{i=0}
    \Big(\begin{matrix}
            \mII \\i
        \end{matrix}\Big)
    f_{\mII,i}\,\ptrans(\mII,i),
\end{equation}
where $m_\mathrm{I}$ ($m_\mathrm{II}$) is the number of \typei{} (\typeii{}) nodes,
$f_{\mII,i}$ is the secret key fraction, $\ptrans(\mII,i)$ is the probability of successful transmission with 1-erasure errors in $i$ distinct \typeii{} nodes (see \methodsref{methods:subsec:repeater-performance}), and
\begin{equation}\label{eqn:processing-time}
    \tau_\mathrm{tot}=
    \tau_\mathrm{tree}+14\tau_\mathrm{ss}+26\tau_\mathrm{tele}+8\tau_\mathrm{meas},
\end{equation}
is the total processing time of each \typeii{} node for 1 logical qubit.
Here, $\tau_\mathrm{ss}$ is the local spin-spin two-qubit gate time, $\tau_\mathrm{tele}$ is the teleported two-qubit gate time, and $\tau_\mathrm{meas}$ is the spin readout time. We arrived at \cref{eqn:processing-time} by considering the longest possible time taken by the fault-tolerant error correction protocol since that will become the processing bottleneck of the network (see \supplmat{}).
We assume that the readout of the ancilla and flag qubit can be done simultaneously.
Finally, a new tree-cluster has to be generated in the \typeii{} node with tree vector $\vec{t}=[b_0,b_1,\ldots,b_d]$ which takes time $\tau_\mathrm{tree}$ and is estimated as \cite{Borregaard2020}
\begin{align}\label{eqn:tau_tree}
    \tau_\mathrm{tree}
     & \approx
    b_0
    \Big[
        100+b_1(1+b_2(1+\cdots b_{d-1}(1+b_d)\cdots))
    \Big]\tau_\mathrm{ph}\nonumber \\
     & +
    b_0
    \Big[
    3+b_1(1+b_2(1+\cdots b_{d-2}(1+b_{d-1})\cdots))
    \Big]\tau_\mathrm{ss},
\end{align}
where $\tau_\mathrm{ph}$ is the emission time of a single photonic qubit and the emission time of the first-level photons is assumed to be $100\tau_\mathrm{ph}$.
The longer emission time of the photons means a narrower frequency bandwidth. This is necessary since the first-level photons may participate in a cavity-mediated \CPHASE{}~gate at the next repeater station. If the frequency bandwidth of the incoming photon is not narrow compared to the cavity (Purcell) enhanced linewidth of the emitter, the gate operation will be imperfect. Assuming an emission time of $100\tau_\mathrm{ph}$ will ensure that these imperfections lead to gate errors $\lesssim10^{-4}$~\cite{Borregaard2020}.

In principle, \typei{} nodes could operate faster than \typeii{} nodes since they are only performing loss correction at the tree-code level.
However, \typei{} nodes would have to wait for the \typeii{} nodes to complete their operations before transmitting the tree-encoded qubits. Thus, the operation time of the \typeii{} nodes becomes the bottleneck that sets the repetition rate of the repeater network and faster \typeii{} nodes are not required. As a consequence, the hardware architecture at \typei{} and \textsc{ii} nodes can be quite similar, facilitating the large scale building of such nodes.

\section{Results}\label{sec:results}
\begin{figure}[t!]
    \centering
\clearpage{}%
\begin{tikzpicture}
    \begin{groupplot}[
            group style={
                    group size=1 by 2,
                    x descriptions at=edge bottom,
                    vertical sep=0pt,
                },
            xmode=log,
            ymode=log,
            xlabel={$L_\mathrm{tot}$ [km]},
            xmin=10^2,
            xmax=10^4,
            grid=none,
            auto title,
        ]
        \nextgroupplot[
            ymin=8*10^1,
            ymax=2*10^5,
            legend style={at={(0.03,0.03)},anchor=south west},
            ylabel={$S\!K\!R$ [Hz]},
        ]
        \addlegendimage{empty legend}
        \addplot [mybla,mark=x,mark size=2.5pt,thick,line cap=round,line join=round] table [x index=0, y expr=\thisrowno{1}, header=true, col sep=comma] {assets/csv/21_points_ea49e5c8/SKR_kappa0001/SKR_reencodingError0.0001_kappa0001.csv};
        \addplot [myblu,mark=o,mark size=2.0pt,thick,line cap=round,line join=round] table [x index=0, y expr=\thisrowno{1}, header=true, col sep=comma] {assets/csv/21_points_ea49e5c8/SKR_kappa0001/SKR_reencodingError0.0003_kappa0001.csv};
        \addplot [myred,mark=diamond,mark size=2.75pt,thick,line cap=round,line join=round] table [x index=0, y expr=\thisrowno{1}, header=true, col sep=comma] {assets/csv/21_points_ea49e5c8/SKR_kappa0001/SKR_reencodingError0.0005_kappa0001.csv};
        \addplot [mypur,mark=square,mark size=1.7pt,thick,line cap=round,line join=round] table [x index=0, y expr=\thisrowno{1}, header=true, col sep=comma] {assets/csv/21_points_ea49e5c8/SKR_kappa0001/SKR_reencodingError0.001_kappa0001.csv};
        \addplot [myora,mark=triangle,mark size=2.5pt,thick,line cap=round,line join=round] table [x index=0, y expr=\thisrowno{1}, header=true, col sep=comma] {assets/csv/21_points_ea49e5c8/SKR_kappa0001/SKR_reencodingError0.002_kappa0001.csv};

        \addplot [mybla, thick, dashed, line cap=round,line join=round] table [x index=0, y expr=\thisrowno{1}*5, header=true, col sep=comma] {assets/csv/homogeneous_SKR_reencodingError0.0001.csv};
        \addplot [myblu, thick, dashed, line cap=round,line join=round] table [x index=0, y expr=\thisrowno{1}*5, header=true, col sep=comma] {assets/csv/homogeneous_SKR_reencodingError0.0003.csv};
        \addplot [myred, thick, dashed, line cap=round,line join=round] table [x index=0, y expr=\thisrowno{1}*5, header=true, col sep=comma] {assets/csv/homogeneous_SKR_reencodingError0.0005.csv};
        \addplot [mypur, thick, dashed, line cap=round,line join=round] table [x index=0, y expr=\thisrowno{1}*5, header=true, col sep=comma] {assets/csv/homogeneous_SKR_reencodingError0.001.csv};
        \addplot [myora, thick, dashed, line cap=round,line join=round] table [x index=0, y expr=\thisrowno{1}*5, header=true, col sep=comma] {assets/csv/homogeneous_SKR_reencodingError0.002.csv};

        \addlegendentry{\hspace{-.6cm}$\epsilon_\mathrm{r}$}
        \addlegendentry{$0.1\text{\textperthousand}$}
        \addlegendentry{$0.3\text{\textperthousand}$}
        \addlegendentry{$0.5\text{\textperthousand}$}
        \addlegendentry{$0.1\%$}
        \addlegendentry{$0.2\%$}

        \nextgroupplot[
            ymin=10^5,
            ymax=2*10^8,
            ylabel={Cost},
        ]
        \addplot [mybla,mark=x,mark size=2.5pt, thick, line cap=round,line join=round] table [x index=0, y expr=\thisrowno{1}, header=true, col sep=comma] {assets/csv/21_points_ea49e5c8/cost_kappa0001/cost_reencodingError0.0001_kappa0001.csv};
        \addplot [myblu,mark=o,mark size=2.0pt, thick, line cap=round,line join=round] table [x index=0, y expr=\thisrowno{1}, header=true, col sep=comma] {assets/csv/21_points_ea49e5c8/cost_kappa0001/cost_reencodingError0.0003_kappa0001.csv};
        \addplot [myred,mark=diamond, mark size=2.75pt, thick, line cap=round,line join=round] table [x index=0, y expr=\thisrowno{1}, header=true, col sep=comma] {assets/csv/21_points_ea49e5c8/cost_kappa0001/cost_reencodingError0.0005_kappa0001.csv};
        \addplot [mypur,mark=square,mark size=1.7pt, thick, line cap=round,line join=round] table [x index=0, y expr=\thisrowno{1}, header=true, col sep=comma] {assets/csv/21_points_ea49e5c8/cost_kappa0001/cost_reencodingError0.001_kappa0001.csv};
        \addplot [myora,mark=triangle,mark size=2.5pt, thick, line cap=round,line join=round] table [x index=0, y expr=\thisrowno{1}, header=true, col sep=comma] {assets/csv/21_points_ea49e5c8/cost_kappa0001/cost_reencodingError0.002_kappa0001.csv};

        \addplot [mybla, thick, dashed, line cap=round,line join=round] table [x index=0, y expr=\thisrowno{1}*5, header=true, col sep=comma] {assets/csv/homogeneous_cost_reencodingError0.0001.csv};
        \addplot [myblu, thick, dashed, line cap=round,line join=round] table [x index=0, y expr=\thisrowno{1}*5, header=true, col sep=comma] {assets/csv/homogeneous_cost_reencodingError0.0003.csv};
        \addplot [myred, thick, dashed, line cap=round,line join=round] table [x index=0, y expr=\thisrowno{1}*5, header=true, col sep=comma] {assets/csv/homogeneous_cost_reencodingError0.0005.csv};
        \addplot [mypur, thick, dashed, line cap=round,line join=round] table [x index=0, y expr=\thisrowno{1}*5, header=true, col sep=comma] {assets/csv/homogeneous_cost_reencodingError0.001.csv};
        \addplot [myora, thick, dashed, line cap=round,line join=round] table [x index=0, y expr=\thisrowno{1}*5, header=true, col sep=comma] {assets/csv/homogeneous_cost_reencodingError0.002.csv};
    \end{groupplot}
\end{tikzpicture}
\clearpage{}%

    \caption{
        (a) The secret key rate $\SKR$ corresponding to
        (b)~the minimized cost function $C_\mathrm{min}$ as a function of the distance $L_\mathrm{tot}$ for various re-encoding error probabilities $\epsilonr$ with fixed repeater relative weight $\kappa=1$.
        For comparison, the secret key rates and the costs of the homogeneous repeater network from Ref.~\cite{Borregaard2020} multiplied by a factor of 5 are included as dashed lines as well.
        The factor of 5 is included to compensate for the fact that the concatenated repeater is similar to 5 parallel homogeneous repeater networks.
        The cost function of the homogeneous repeater is the same as \cref{eqn:cost-function} with $\mII=0$, and was minimized independently.
    }
    \label{fig:skr-cost-function-plot}
\end{figure}
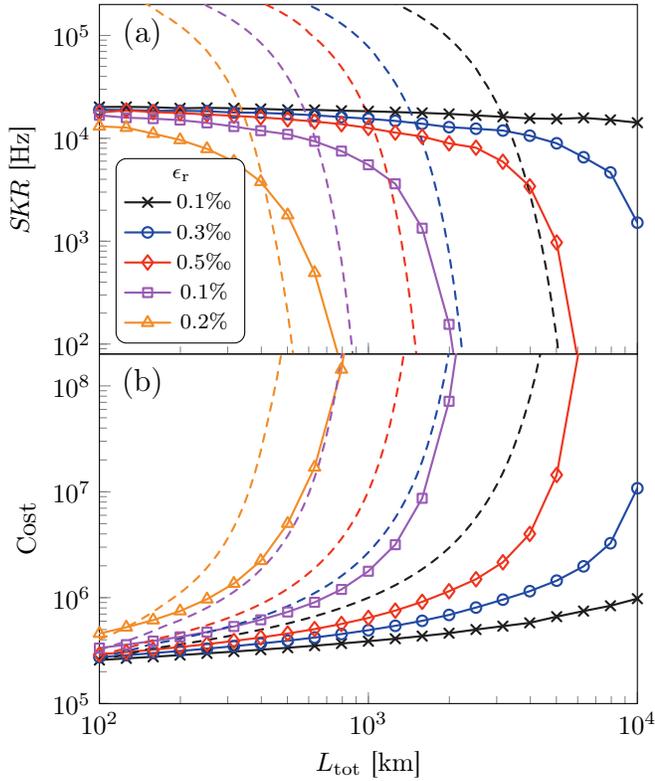

\begin{table}[t!]
    \centering
    \begin{tabular}{l@{\hspace{0.4em}}l@{ }c@{\hspace{2.6em}}c} \toprule[1.3pt]
        \multicolumn{2}{l}{Quantity}           &                      & Value                                                 \\ \midrule \\ [-2.0ex]
        Spin-spin gate time,                   & $\tau_\mathrm{ss}$   & $\scriptscriptstyle\mathbin{\Diamond}$ & $100$ ns     \\
        Single photon emission time,           & $\tau_\mathrm{ph}$   & $\scriptscriptstyle\mathbin{\Diamond}$ & $1$ ns       \\
        Spin readout time,                     & $\tau_\mathrm{meas}$ &                                        & $1$ $\upmu$s \\
        Teleported two-qubit gate time,        & $\tau_\mathrm{tele}$ &                                        & $1$ $\upmu$s \\
        Optical fiber's attenuation length,    & $L_\mathrm{att}$     & $\scriptscriptstyle\mathbin{\Diamond}$ & $20$ km      \\
        Effective photon-detection efficiency, & $\etad$              & $\scriptscriptstyle\mathbin{\Diamond}$ & $0.95$       \\ \bottomrule[1.3pt]
    \end{tabular}
    \caption{Values of the quantities used in the minimization of the cost function shown in \cref{eqn:cost-function}. The symbol $\scriptscriptstyle\mathbin{\Diamond}$ denotes that the values were also used in Ref.~\cite{Borregaard2020}. The effective photon-detection efficiency, $\etad$ is the combined efficiency of in/out coupling of the photon, frequency conversion and the efficiency of the photon detectors. }
    \label{tab:contants}
\end{table}

To find the optimal configuration of the repeater network, that is to maximize the secret key rate while using as little physical resources as possible, we perform a numerical minimization of a dimensionless cost function for a specific total distance
\begin{equation}\label{eqn:cost-function}
    C=\SKR^{-1}\frac{L_\mathrm{att}}{\tau_\mathrm{ph}L_\mathrm{tot}}(m_\mathrm{I}+\kappa m_\mathrm{II}),
\end{equation}
over the parameters $L_0$, $\mII$, and $\vec{t}$ with constraints discussed in \methodsref{methods:subsec:numerical-minimization}. Note that for simplicity, we assume a uniform interspersing of \typei{} and \typeii{} nodes in the repeater network. For more details on how the repeaters are interspersed, refer to the \supplmat{}.
The coefficient $\kappa$ is the relative cost of a \typeii{} to a \typei{} node. In a repeater network with multiple types of nodes, one type of node may require more functionality (as is the case here) and thus be more expensive than another.
For instance, if $\kappa=1$, then the cost of a \typei{} node is the same as that of a \typeii{} node.

To quantify the cost per unit length, we divide the number of nodes in the cost function with the total distance between the start and end node $L_\mathrm{tot}$.
This is expressed in units of the attenuation length $L_\mathrm{att}$ to make the cost function dimensionless.
We also quantify the cost per unit time by including the inverse of the secret key rate ($\SKR$) which is in unit Hertz.
Note that the secret key rate depends on the photon emission time $\tau_\mathrm{ph}$, therefore its presence in the denominator in \cref{eqn:cost-function} serves to make the cost function dimensionless. It also means that we are expressing the cost per unit time in units of the photon emission time.
The values for the constants used are shown in \cref{tab:contants}.

The results of the optimizations for different error rates are shown in \cref{fig:skr-cost-function-plot} for $\kappa=1$.
For short distances, the homogeneous repeater scheme from Ref.~\cite{Borregaard2020} (dashed lines) is superior since it does not possess the time overhead of error correction that comes with \typeii{} nodes and thus enables a higher repetition rate.
For longer total distances, however, the secret key rate of the concatenated repeater protocol (solid lines with markers) greatly surpasses that of the homogeneous repeater protocol of Ref.~\cite{Borregaard2020} due to the added protection from the outer 5-qubit code.
For instance, for $\epsilonr=0.1\%$ (purple line), the secret key rate for the concatenated repeater is $\sim\!\!5.5$~kHz at~$10^3$~km, while for the homogeneous counterpart (dashed line), the rate at the same total distance is less than 1~Hz. Furthermore, despite the higher secret key rates of the homogeneous repeater scheme, we see that from \figref{fig:skr-cost-function-plot}{b} that its cost is significantly higher than that of the concatenated repeater protocol. This means that the secret key rate itself is not the only relevant indicator of overall performance, but the cost associated with building the proposed network needs to be considered as well.
Note that even if the cost function is ignored and if the repeater network is configured by purely maximizing the secret key rate, the gain in the secret key rate would be minimal with the tradeoff being a significantly increased cost (see \supplmat{}).

\begin{figure}[t!]
    \centering
\clearpage{}%
\begin{tikzpicture}
    \begin{groupplot}[
            group style={
                    group size=1 by 2,
                    x descriptions at=edge bottom,
                    vertical sep=0pt,
                },
            xmode=log,
            xlabel={$L_\mathrm{tot}$ [km]},
            xmin=10^2,
            xmax=10^4,
            grid=none,
            auto title,
        ]
        \nextgroupplot[
            ymode=linear,
            ymin=0.5,
            ymax=4.5,
            ytick={1,2,3,4},
            legend style={at={(0.99,0.98)},anchor=north east},
            ylabel={$L_0$ [km]},
        ]
        \addlegendimage{empty legend}
        \addplot [mybla,mark=x,mark size=2.5pt, thick, line cap=round,line join=round] table [x index=0, y expr=\thisrowno{1}, header=true, col sep=comma] {assets/csv/21_points_ea49e5c8/L0_kappa0001/L0_reencodingError0.0001_kappa0001.csv};
        \addplot [myblu,mark=o,mark size=2.0pt, thick, line cap=round,line join=round] table [x index=0, y expr=\thisrowno{1}, header=true, col sep=comma] {assets/csv/21_points_ea49e5c8/L0_kappa0001/L0_reencodingError0.0003_kappa0001.csv};
        \addplot [myred,mark=diamond, mark size=2.75pt, thick, line cap=round,line join=round] table [x index=0, y expr=\thisrowno{1}, header=true, col sep=comma] {assets/csv/21_points_ea49e5c8/L0_kappa0001/L0_reencodingError0.0005_kappa0001.csv};
        \addplot [mypur,mark=square,mark size=1.7pt, thick, line cap=round,line join=round] table [x index=0, y expr=\thisrowno{1}, header=true, col sep=comma] {assets/csv/21_points_ea49e5c8/L0_kappa0001/L0_reencodingError0.001_kappa0001.csv};
        \addplot [myora,mark=triangle,mark size=2.5pt, thick, line cap=round,line join=round] table [x index=0, y expr=\thisrowno{1}, header=true, col sep=comma] {assets/csv/21_points_ea49e5c8/L0_kappa0001/L0_reencodingError0.002_kappa0001.csv};
        \addlegendentry{\hspace{-.6cm}$\epsilon_\mathrm{r}$}
        \addlegendentry{$0.1\text{\textperthousand}$}
        \addlegendentry{$0.3\text{\textperthousand}$}
        \addlegendentry{$0.5\text{\textperthousand}$}
        \addlegendentry{$0.1\%$}
        \addlegendentry{$0.2\%$}
        \nextgroupplot[
            ymode=linear,
            ymin=0,
            ymax=9,
            ylabel={$\mI:\mII$},
            yticklabel style={ %
                    /pgf/number format/fixed,
                    /pgf/number format/precision=2,
                },
        ]
        \addplot [mybla,mark=x,mark size=2.5pt, thick, line cap=round,line join=round] table [x index=0, y expr=1/(\thisrowno{1}), header=true, col sep=comma] {assets/csv/21_points_ea49e5c8/mII-to-mI_kappa0001/mII-to-mI-ratio_reencodingError0.0001_kappa0001.csv};

        \addplot [myblu,mark=o,mark size=2.0pt, thick, line cap=round,line join=round] table [x index=0, y expr = 1/(\thisrowno{1}), header=true, col sep=comma] {assets/csv/21_points_ea49e5c8/mII-to-mI_kappa0001/mII-to-mI-ratio_reencodingError0.0003_kappa0001.csv};
        \addplot [myred,mark=diamond, mark size=2.75pt, thick, line cap=round,line join=round] table [x index=0, y expr = 1/(\thisrowno{1}), header=true, col sep=comma] {assets/csv/21_points_ea49e5c8/mII-to-mI_kappa0001/mII-to-mI-ratio_reencodingError0.0005_kappa0001.csv};
        \addplot [mypur,mark=square,mark size=1.7pt, thick, line cap=round,line join=round] table [x index=0, y expr = 1/(\thisrowno{1}), header=true, col sep=comma] {assets/csv/21_points_ea49e5c8/mII-to-mI_kappa0001/mII-to-mI-ratio_reencodingError0.001_kappa0001.csv};
        \addplot [myora,mark=triangle,mark size=2.5pt, thick, line cap=round,line join=round] table [x index=0, y expr = 1/(\thisrowno{1}), header=true, col sep=comma] {assets/csv/21_points_ea49e5c8/mII-to-mI_kappa0001/mII-to-mI-ratio_reencodingError0.002_kappa0001.csv};
    \end{groupplot}
\end{tikzpicture}
\clearpage{}%

    \caption{
        (a) The inter-repeater distance $L_0$, and
        (b)~the ratio of number of \typei{} to \typeii{} nodes $\mI:\mII$ as a function of total distance corresponding to \cref{fig:skr-cost-function-plot}.
    }
    \label{fig:parameters-plot}
\end{figure}
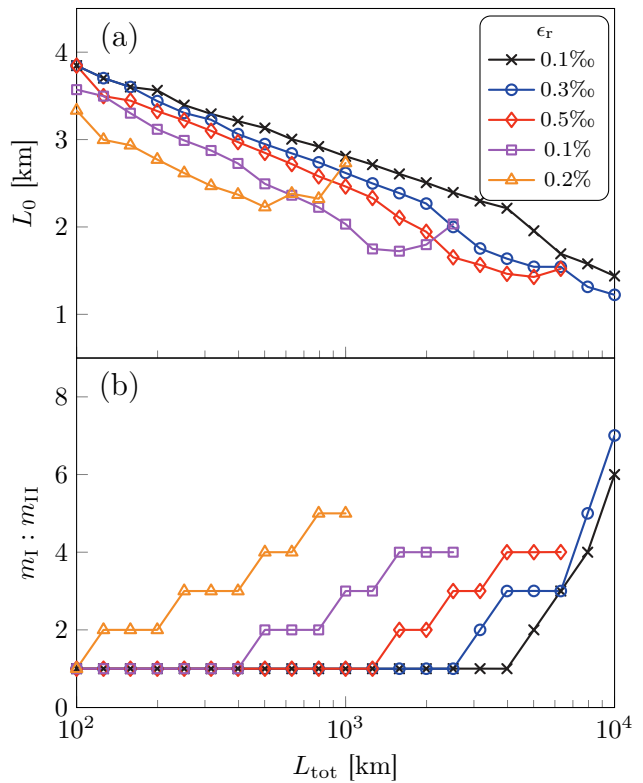

The accompanying results in \figref{fig:parameters-plot}{a} reveal the optimal inter-repeater distances required to achieve the best secret key rates whilst maintaining minimum cost.
When considering higher re-encoding error probabilities $\epsilonr$ for fixed total distances, the optimum inter-repeater distance~$L_0$~decreases to minimize the loss error to counter the increasing re-encoding error rate.
From \figref{fig:parameters-plot}{b}, we see that the configuration with the highest re-encoding error has the lowest proportion of \typeii{} nodes in the network, even though it has the lowest inter-repeater distance. This is because the re-encoding error rate scales with the error rate of the two-qubit gates in the \typeii{} nodes as explained in \cref{subsec:re-encoding}. Therefore, adding more \typeii{} nodes no longer entails better error suppression, but instead introduces more error in the network.
Hence, we find an asymmetry between the number of \typei{} and \typeii{} nodes as the ratio of $\mI$ to $\mII$ deviates from unity. This result demonstrates that even when we consider the cost of \typei{} and \typeii{} nodes to be equal, i.e., $\kappa=1$, the asymmetry between loss and operation errors in the repeater network is still present.
Note that it is in principle possible for $\kappa$ to have values below unity. In our calculation, however, because of the constraint on the $\mII$ set, the minimum value that the ratio of $\mI$ to $\mII$ can attain is unity.

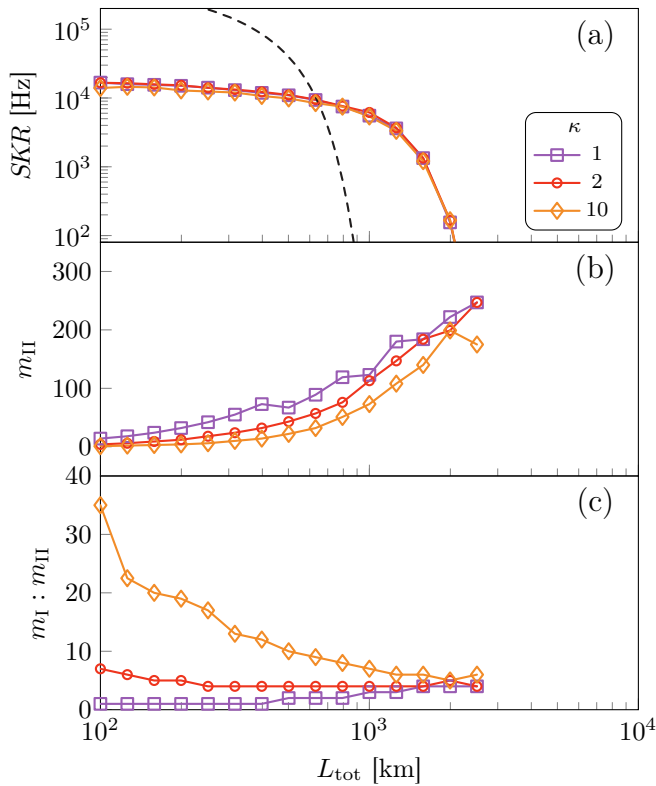
\begin{figure}[t!]
    \centering
\clearpage{}%
\begin{tikzpicture}
    \begin{groupplot}[
            group style={
                    group size=1 by 3,
                    x descriptions at=edge bottom,
                    vertical sep=0pt,
                },
            xmode=log,
            xlabel={$L_\mathrm{tot}$ [km]},
            xmin=10^2,
            xmax=10^4,
            grid=none,
            auto title,
            height=4.69cm,
            title style={at={(0.92,0.71)},font=\large},
        ]
        \nextgroupplot[
            ymode=log,
            ymin=8*10^1,
            ymax=2*10^5,
            legend style={at={(0.97,0.05)},anchor=south east},
            ylabel={$S\!K\!R$ [Hz]},
        ]
        \addlegendimage{empty legend}
        \addplot+ [mypur,mark=square,mark size=2.2pt, thick, line cap=round,line join=round] table [x index=0, y expr=\thisrowno{1}, header=true, col sep=comma] {assets/csv/21_points_ea49e5c8/SKR_kappa0001/SKR_reencodingError0.001_kappa0001.csv};
        \addplot+ [myred,mark=o,mark size=1.7pt, thick, line cap=round,line join=round,mark options=solid] table [x index=0, y expr=\thisrowno{1}, header=true, col sep=comma] {assets/csv/21_points_88ed70e9/SKR_kappa0002/SKR_reencodingError0.001_kappa0002.csv};
        \addplot+ [myora,mark=diamond,mark size=3.05pt, thick, line cap=round,line join=round] table [x index=0, y expr=\thisrowno{1}, header=true, col sep=comma] {assets/csv/21_points_b5509d17/SKR_kappa0010/SKR_reencodingError0.001_kappa0010.csv};

        \addplot [mybla, thick, dashed, line cap=round,line join=round] table [x index=0, y expr=\thisrowno{1}*5, header=true, col sep=comma] {assets/csv/homogeneous_SKR_reencodingError0.001.csv};

        \addlegendentry{\hspace{-.6cm}$\kappa$}
        \addlegendentry{$1$}
        \addlegendentry{$2$}
        \addlegendentry{$10$}
        \nextgroupplot[
            ymode=linear,
            ymin=-50,
            ymax=350,
            ytick={0,100,200,300},
            ylabel={$\mII$},
        ]
        \addplot+ [mypur,mark=square,mark size=2.2pt, thick, line cap=round,line join=round] table [x index=0, y expr=\thisrowno{1}, col sep=comma] {assets/csv/21_points_ea49e5c8/mII_kappa0001/mII_reencodingError0.001_kappa0001.csv};
        \addplot+ [myred,mark=o,mark size=1.7pt, thick, line cap=round,line join=round] table [x index=0, y expr=\thisrowno{1}, col sep=comma] {assets/csv/21_points_88ed70e9/mII_kappa0002/mII_reencodingError0.001_kappa0002.csv};
        \addplot+ [myora,mark=diamond,mark size=3.05pt, thick, line cap=round,line join=round] table [x index=0, y expr=\thisrowno{1}, col sep=comma] {assets/csv/21_points_b5509d17/mII_kappa0010/mII_reencodingError0.001_kappa0010.csv};
        \nextgroupplot[
            ymode=linear,
            ymin=0,
            ymax=40,
            ylabel={$\mI:\mII$},
        ]
        \addplot [mypur,mark=square,mark size=2.2pt, thick, line cap=round,line join=round] table [x index=0, y expr = 1/(\thisrowno{1}), header=true, col sep=comma] {assets/csv/21_points_ea49e5c8/mII-to-mI_kappa0001/mII-to-mI-ratio_reencodingError0.001_kappa0001.csv};
        \addplot [myred,mark=o,mark size=1.7pt, thick, line cap=round,line join=round] table [x index=0, y expr = 1/(\thisrowno{1}), header=true, col sep=comma] {assets/csv/21_points_88ed70e9/mII-to-mI_kappa0002/mII-to-mI-ratio_reencodingError0.001_kappa0002.csv};
        \addplot [myora,mark=diamond,mark size=3.05pt, thick, line cap=round,line join=round] table [x index=0, y expr = 1/(\thisrowno{1}), header=true, col sep=comma] {assets/csv/21_points_b5509d17/mII-to-mI_kappa0010/mII-to-mI-ratio_reencodingError0.001_kappa0010.csv};
    \end{groupplot}
\end{tikzpicture}
\clearpage{}%

    \caption{
        (a) The secret key rate $\SKR$ corresponding to the minimized cost function $C_\mathrm{min}$ as a function of the distance $L_\mathrm{tot}$ for various relative node weights $\kappa$ with fixed re-encoding error probability $\epsilonr=0.1\%$.
        For comparison, the secret key rate of the homogeneous repeater network multiplied by a factor of 5 from Ref.~\cite{Borregaard2020} is included as dashed line optimized with respect to the cost function shown in \cref{eqn:cost-function} with fixed $\mII=0$.
        (b)~The corresponding number of \typeii{} nodes in the network and
        (c)~the corresponding ratio of number of \typei{} to \typeii{} nodes as a function of total distance.
    }
    \label{fig:varying-kappa}
\end{figure}

To demonstrate how the value of $\kappa>1$ influences the cost and secret key rate when considering \typeii{} nodes that are more expensive than \typei{} nodes, we optimize with $\kappa\in\{1,2,10\}$ for fixed re-encoding error rate $\epsilonr=0.1\%$ in \cref{fig:varying-kappa}.
As the value of $\kappa$ increases, we see that the secret key rate is only minimally affected as shown in \figref{fig:varying-kappa}{a} whilst the number of \typeii{} nodes in the network is significantly decreased as shown in \figref{fig:varying-kappa}{b}. From \figref{fig:varying-kappa}{c}, we infer that this is followed by an increase in the number of \typei{} nodes in the network. This is expected as \typeii{} nodes are more costly.
Additionally, \figref{fig:varying-kappa}{c} shows how the ratio of \typei{} to \typeii{} nodes changes as function of the distance.
For high-cost \typeii{} nodes ($\kappa=10$) the ratio decreases with distance.
This can be understood from the accumulation of re-encoding errors in the network.
For small distances, the accumulated error is relatively small, and it is therefore advantageous to employ few \typeii{} nodes.
For larger distances, the accumulated error necessitates more \typeii{} nodes despite their higher cost.
On the other hand, we observe that the ratio of \typei{} to \typeii{} nodes increases with distance even for a low cost of \typeii{} nodes ($\kappa=1$).
We attribute this to the fact that placing too many \typeii{} nodes in the network introduces more operational errors than it can suppress.

\section{Discussion}\label{sec:conclusion}
In conclusion, we have outlined how a resource-efficient DV one-way quantum repeater can be constructed by concatenating a flag-based 5-qubit~code with a loss-tolerant tree-cluster code.
The high loss tolerance and conceptually straightforward encoding/decoding of the tree-cluster code makes it suitable as an inner code that protects against photon loss in transmission between repeater nodes.
An extra layer of protection is provided by the outer 5-qubit~code which effectively suppresses operational errors that accumulate due to the non-fault-tolerant nature of the tree-cluster code against Pauli errors.
As a result, the code-concatenated repeater is able to bridge distances of several thousand kilometers for re-encoding error probabilities as high as $\sim\!0.1\%$.

The code concatenation allows for an architecture tailored to the asymmetry between loss and operational errors that is characteristic for quantum communication.
Specifically, we have shown how the optimizations can include relative cost of the different repeater nodes to arrive at an optimized architecture with many relatively cheap \typei{} nodes that only correct errors due to transmission loss, but far fewer expensive \typeii{} nodes that also correct operational errors.
In particular, we found that such an optimization could be done with a minimal effect on the optimal secret key rate of the repeater network.

This represents major advantage because it achieves fault-tolerant operation with a modest resource overhead and allows for long-distance communication with error rates roughly an order of magnitude larger compared to the non-fault-tolerant scheme of Ref.~\cite{Borregaard2020}.
Furthermore, the DV nature of our protocol circumvents the highly challenging generation of optical GKP states, making our proposal applicable to a wide range of experimental systems, which are predominantly qubit-based.
In the supplemental material, we provide further details on the comparison between our work and previous repeater protocols outlining how our scheme has a minimized number of spin qubits per repeater node.
Due to the flexibility of our architecture, it is possible to use other quantum error correcting codes such as the \textnkd{7,1,3} and \textnkd{9,1,3} codes or the \textnkd{4,2,2} quantum error detecting code with minimal modification.
We leave it up to future work to investigate of such approaches could lead to better results in terms of the choice of code, network performance, and resource requirement.

We have outlined a possible modular implementation of the repeater based on few-qubit processors with an efficient spin-photon interface through a cavity-coupled quantum emitter.
Solid-state systems such as group-IV defect centers in diamond \cite{Nguyen2019,Bhaskar2020,Rugar2020} are promising hardware candidates having demonstrated both efficient coupling to nanophotonic cavities~\cite{Bhaskar2020} and access to small-scale qubit registers through coupling to nearby nuclear spins~\cite{Nguyen2019}.
To implement both \typei{} and \typeii{} nodes, only 5 cavity-emitter systems are required. Importantly, it is sufficient for each of these systems to have just 1 cavity-coupled emitter spin with direct, e.g., magnetic, coupling to 4 nearby nuclear spins.
Operations between the processors, which are needed for the \typeii{} nodes can be achieved through teleported gates facilitated by photon-mediated interactions between the quantum-emitter spins.

\section{Methods}\label{sec:methods}
\subsection{Error correction}\label{methods:subsec:error-correction}
\subsubsection{Fault-tolerance}
\begin{figure}[b!]
    \centering
    \includegraphics[width=0.9\linewidth]{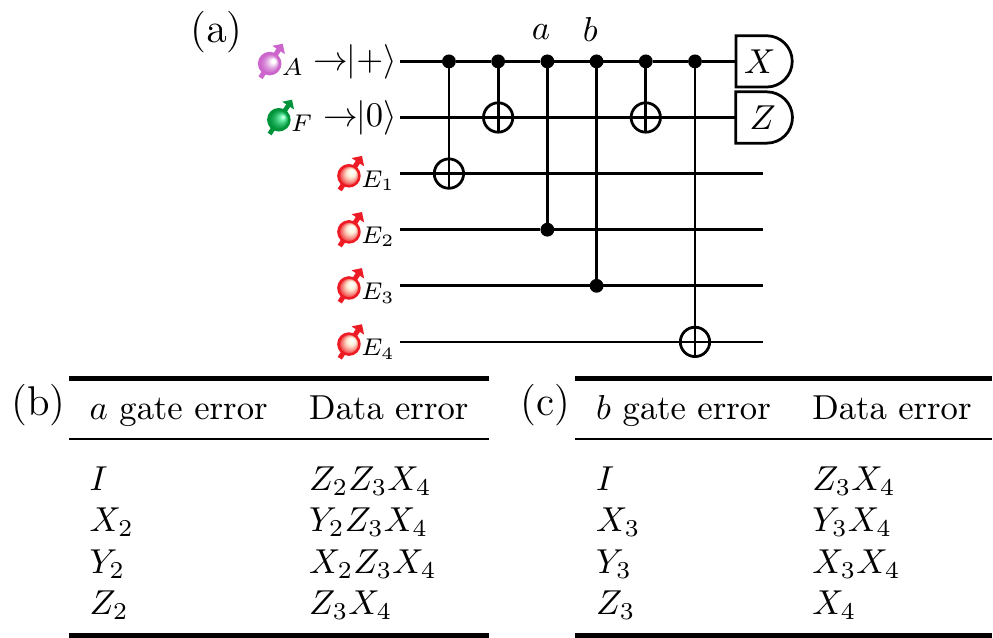}
    \caption{(a) Circuit diagram of the error propagation in a subsection of the 5-qubit code circuit. (b) The errors induce by gate $a$ if it was triggered by an error on the ancilla qubit and (c) similarly for gate $b$. Figure adapted from Fig.~1 in Ref.~\cite{Chao2018}.}
    \label{fig:flag_error_propagate}
\end{figure}

In a \typeii{} node, an additional flag qubit in a \typeii{} node is used to perform error correction, with a circuit shown in \cref{fig:flag_qcircuit}. This additional qubit aids in suppressing the otherwise undetectable physical errors propagated by faulty gates in the stabilizer operations \cite{Chao2018}.
In \cref{fig:flag_error_propagate}, we consider a subsystem of \cref{fig:flag_qcircuit} as an example and illustrate how an error that occurred right before a two-qubit gate during the stabilizer operation could result in logical errors on the data qubits due to the error being propagated by the two-qubit gate onto the data qubits. Then, we consider additional errors induced by the noisy two-qubit gate on the target qubit.
With one additional flag qubit prepared in state $\ket{0}$, the propagated error on the data qubits would then be detected because the flag qubit would be measured as being in state $\ket{1}$.
When the flag qubit is detected as being in state $\ket{1}$, we switch over to the unflagged circuit and apply the Pauli operations according to the syndromes measured during this unflagged circuit to revert the propagated error. An example is shown in \figref{fig:flag_error_propagate}{b-c}.
For the complete fault-tolerant error-correction protocol, see the \supplmat{}.

\subsubsection{Erasure-error correction}\label{methods:subsubsec:erasure-error-correction}
In the event of a 1-erasure error being heralded at a \typeii{} node, we initialize the spin in the node corresponding to the lost tree to state $|{0}\rangle$, while the rest of the intact trees are decoded into spins in the other nodes as usual.
The 5-qubit code stabilizer operations are then performed on the 5 qubits, projecting the mixed state back into the logical codespace of the 5-qubit code, up to some Pauli corrections on the 5 qubits.
For example, if the 5$^\mathrm{th}$ qubit that was part of the encoded 5-qubit logical state $|\psi_L\rangle$ was lost, then the original encoded state would be restored according to the Pauli corrections in \cref{tab:erasure-correction-projection}.
The procedure is similar for correcting a 2-erasure error, however, we have not included corrections of 2-erasure errors in our optimizations since they occur very rarely compared to 1-erasure errors.

To complete the error correction look-up table, we considered the cases in which additional Pauli errors also occurred on other non-lost qubits. The resulting look-up table is shown in \cref{tab:1erasure-lut}. Note that this look-up table is not unique, i.e., multiple distinct errors could lead to the same syndrome. For instance, if the $1{\mathrm{st}}$ qubit is lost, the syndrome $\synd{-}{+}{+}{+}$ corresponds to the cases where either a $X_2$, $Z_3$, $Z_4$, or $X_5$ error also occurred. Since they are indistinguishable due to the shared syndrome, we choose to only correct for $X_2$.
Randomly choosing one out of the four possible errors to correct does not affect the suppression of error because we are considering the depolarizing noise model where every Pauli error occurs with the same probability.
However, if one is considering a biased noise model, it is wise to choose which case to correct in order to maximize the error suppression.

Note that for erasure correction, we do not use the error correction protocol with the flag qubit since we found that it does not improve the fidelity of the error corrected state compared to not using the flag qubit. This is due to the fact that the effectiveness of the flag qubit protocol relies on the initial 5-qubit logical state having at most a Pauli error on 1 physical qubit, while an erasure error effectively induces a correlated error, i.e., it is not merely a single qubit Pauli error.

\begin{table}[h!]
    \begin{tabular}{c@{\hskip 3em}c} \toprule[1.3pt]
        Syndrome            & Projected state        \\\midrule \\ [-2.0ex]
        $\synd{+}{+}{+}{+}$ & $|\psi_L\rangle$       \\
        $\synd{+}{+}{-}{-}$ & $X_5|\psi_L\rangle$    \\
        $\synd{+}{-}{+}{+}$ & $Z_5|\psi_L\rangle$    \\
        $\synd{+}{-}{-}{-}$ & $X_5Z_5|\psi_L\rangle$ \\ \bottomrule[1.3pt]
    \end{tabular}
    \caption{The syndrome observed from reading out the ancilla qubits and the corresponding restored encoded logical states assuming the 5$^\mathrm{th}$ qubit was lost.}
    \label{tab:erasure-correction-projection}
\end{table}

\begin{table}[h!]
    \includegraphics[width=0.85\linewidth]{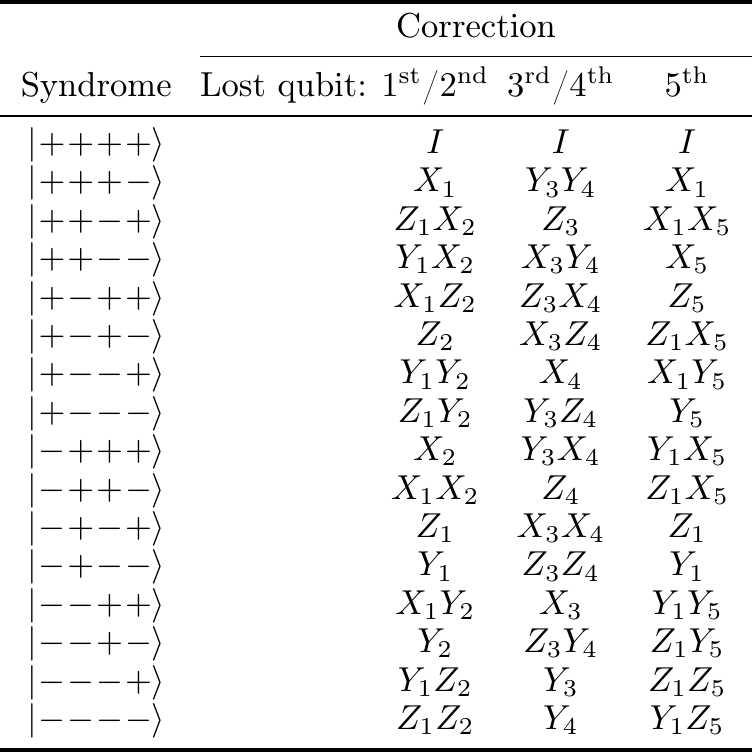}
    \caption{Corrections to the 5 data qubits to project them back into the logical codespace of the 5-qubit~code depending on the ancilla outcomes and which data qubit was lost.}
    \label{tab:1erasure-lut}
\end{table}

\subsection{Error model}\label{methods:subsec:error-model}
We model the noise in the two-qubit gates between two local spins using the two-qubit depolarizing channel
\begin{equation}\label{eqn:two-qubit-depolarizing-noise}
    \Lambda_2(\epsilon)=(1-\epsilon)\varrho+\frac{\epsilon}{15}\sum_{P\in\{I,X,Y,Z\}^{\otimes 2}\setminus{I^{\otimes 2}}}P\varrho P.
\end{equation}
where $\epsilon$ is a general error rate and $\varrho$ is a two-qubit density matrix. In our model, a non-teleported two-qubit gate is followed by $\Lambda_{2}(\epsilon_0)$. Conversely, a teleported two-qubit gate, as shown in \cref{fig:teleported-cnot}, involves 3 two-qubit gates that each has an error of $\epsilon_0$. Thus, we consider that an error of $\Lambda_{2}(3\epsilon_0)$ follows immediately after a teleported two-qubit gate since $\epsilon_0\ll1$.

\begin{figure}[t!]
    \centering
    \includegraphics[width=0.75\linewidth]{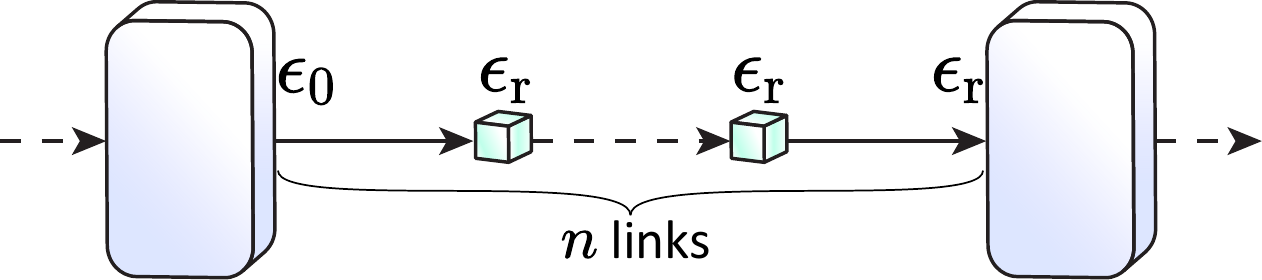}
    \caption{Illustration of the error propagated by the repeater operations between two consecutive \typeii{} nodes with $n$ (parallel) links in between. Only 1 out of 5 of the parallel links is shown for simplicity.}
    \label{fig:etrans}
\end{figure}

Another noise channel which we consider is the single qubit depolarizing channel, which is given by
\begin{equation}\label{eqn:single-qubit-depolarizing-noise}
    \Lambda(\epsilon)=(1-\epsilon)\rho+\frac{\epsilon}{3}\sum_{P\in\{X,Y,Z\}}P\rho P,
\end{equation}
where $\rho$ is a single qubit density matrix. As previously noted in \cref{sec:protocol}, each of the 5 logical qubits at the tree code level in transmission is subjected to such error channel with an error rate of $\epsilontrans$, i.e., $\Lambda(\epsilontrans)$. The expression for the transmission error is given by
\begin{equation}
    \epsilontrans=1-(1-\epsilonr)^{n}(1-\epsilon_0).
\end{equation}
where $\epsilonr$ is the re-encoding error and $n$ is the number of links between consecutive \typeii{} nodes as illustrated in \cref{fig:etrans}. Note that we have parameterized the number of \typei{} nodes as $\mI=\mII(n-1)$.
We consider the error induced by both the tree decoding and encoding step in \typeii{} nodes. The tree encoding step in \typeii{} nodes introduce an error of $\epsilon_0$.
Conversely, the decoding step at a \typei{} and \typeii{} node is similar, hence it introduces an error of $\epsilonr$.
However instead of decoding the tree upon reception into a fresh tree, \typeii{} nodes would perform a \textsc{SWAP} gate, which we assume has an error of $\epsilon_0$, between the decoded qubit from a tree and an auxiliary memory spin in preparation for a teleported two-qubit gate for the 5-qubit code stabilizer operations as detailed in \cref{sec:experimental-considerations}.

Regarding the tree generation scheme from Ref.~\cite{Borregaard2020} which we consider in our work, we note that, in principle, it could lead to correlated errors in the tree cluster states since it involves repeated photon emission from the same emitter with error rate of $\epsilon_0$. This error on the emitter thus might propagate to more than a single photon in the tree.
It is however, outside the scope of this work to consider the effect of such correlated errors. We note that they could potentially be kept small by e.g., intermediate error detection steps in the generation procedure \cite{Stas2022}.
We assume that we are able to suppress the correlated errors, then we can say that each photon in our trees is merely subjected to $\Lambda(\epsilon_0)$.
From our numerical simulation of the tree re-encoding procedure, we find that our model results in a re-encoding error rate of $\epsilonr\approx3\epsilon_0$ at each \typei{} node for the tree sizes considered in this work (see \supplmat{}).
This reflects the fact that despite the tree-clusters consisting of hundreds of photons each subjected to $\Lambda(\epsilon_0)$, the dominant error comes from errors on the first-level photon and spin qubit involved in the re-encoding step.
Errors on the remaining photons, which are measured out, are efficiently suppressed by employing a majority-voting correction strategy.

\subsection{Repeater Performance}\label{methods:subsec:repeater-performance}
The key performance metrics of the repeater network that we consider are the end-to-end transmission probability of the message qubit and the quality of the transmitted qubit. To characterize the end-to-end transmission probability of the message qubit, we start by considering the transmission probability, $\etae$, of a single tree-cluster state with branching vector $\vec{t}=[b_0,b_1,\ldots,b_d]$ between consecutive repeater nodes.
From Ref.~\cite{Varnava2006}, we have that
\begin{equation}\label{eqn:etae_tree_recursive_formula}
    \etae=
    [(1-\mu+\mu R_1)^{b_0}-(\mu R_1)^{b_0}]
    (1-\mu+\mu R_2)^{b_1},
\end{equation}
where $R_k=1-[1-(1-\mu)(1-\mu+\mu R_{k+2})^{b_{k+1}}]^{b_k}$ with $R_{d+1}=0$, $b_{d+1}=0$, and $\mu=1-\eta\etad$.
Here, $\eta=\exp(-L_0/L_\mathrm{att})$ is the transmission probability of a single photon  between repeater nodes with $L_0$ being the inter-node distance and $L_\mathrm{att}$ being the attenuation length of the optical fiber. Furthermore, $\etad$ is the combined efficiency of in/out coupling of the photon, frequency conversion and the photon detection. We assume that efficient frequency conversion to the telecom band is possible such that $L_\mathrm{att}=20$ km.

Without erasure-error correction on the 5-qubit code level, the end-to-end transmission probability of the message qubit would be given by $\etae^{5m_\mathrm{tot}}$ where $m_\mathrm{tot}=\mI+\mII$ is the total number of nodes in the repeater excluding the start node with $\mI$ ($\mII$) being the number of \typei{} (\typeii{}) nodes in the network. Note that we exclude the start node but include the end node when counting $\mII$. Nonetheless, it is advantageous to leverage the outer~5-qubit~code to also correct for erasure errors. Thus, we need to consider how this affects the end-to-end transmission probability of the message qubit.

We choose to treat cases with more than one erasure error, i.e., more than one failed tree encoding, at the same \typeii{} node as a failed transmission for simplicity as mentioned in \cref{methods:subsubsec:erasure-error-correction}.
The total transmission probability in this case can then be expressed as
\begin{equation}
    \sum^{\mII}_{i=0}
    \Big(\begin{matrix}
            \mII \\i
        \end{matrix}\Big)
    \ptrans(\mII,i),
\end{equation}
with
\begin{equation}\label{eqn:ptrans_i}
    \ptrans(\mII,i)=[\etae^{5n}]^{\mII-i}[5\etae^{4n}(1-\etae^n)]^i,\quad \mII\geq i,
\end{equation}
being the probability of successful transmission with 1-erasure errors in $i$ distinct \typeii{} nodes.
Besides the transmission probability we also need to assess the quality of the transmitted qubits.
To this end, we consider a scenario where the transmitted qubits are used to distill a secret key.
In particular, we assume that the six-state protocol \cite{Bruss1998} is employed, which allows us to quantify the quality of the transmitted qubits using its asymptotic secret key fraction. This is given by~\cite{Scarani2009}
\begin{equation}\label{eqn:secret_key_fraction_repeater}
    f_{\mII,i}=\max\!\Big((1-Q)\Big[1-h\Big(\frac{1-3Q/2}{1-Q}\Big)\Big]-h(Q),0\Big),
\end{equation}
where $h(x)=-x\log_2x-(1-x)\log_2(1-x)$ is the binary entropy and
\begin{equation}\label{eqn:qber}
    Q=2\epsiloneff(\mII,i)/3,
\end{equation}
denotes the QBER (QuBit Error Rate) of the fully decoded qubit at the end node. The quantity $\epsiloneff(\mII,i)$ is the effective error rate of the received qubit at the end node given there were $i$ erasure-error occurrences and $\mII$ \typeii{} nodes.

With our error model, we determine $\epsiloneff(\mII,i)$ from numerical simulations of the 5-qubit code throughout the repeater network.
The details of these simulations can be found in the \supplmat{} where we also provide a semi-analytical approximation to $\epsiloneff(\mII,i)$, which matches the numerical results to great precision for the relevant range of effective error rates, i.e., error rate that is less than the threshold QBER of the associated QKD protocol.

\subsection{Numerical minimization}\label{methods:subsec:numerical-minimization}

We minimize the cost function $C$ in \cref{eqn:cost-function} with respect to $L_0$, $\mII$, and $\vec{t}$ with the following constraints: minimum inter-repeater distance $L_0\geq1$~km, the maximum number of \typeii{} nodes never exceed half of the total number of nodes, i.e., $\mII\leq\lfloor m_\mathrm{tot}/2\rfloor$, the maximum photon number~$N_\mathrm{max}=300$ with the total photon number~$N=\sum^{d}_{i=0}\prod^{i}_{j=0}b_j$ and a fixed depth of the tree $d=2$.
Then, the minimization of $C$ can be written as
\begin{align}\label{eqn:cost-function-minimization}
    C_\mathrm{min} & =\min_{L_0,\mII,\vec{t}}C,\;\;\text{subject to: }\nonumber                        \\
                   & \phantom{{}==={}}\text{$L_0\geq1$ km},\nonumber                                   \\
                   & \phantom{{}==={}}\text{$1\leq\mII\leq \lfloor m_\mathrm{tot}/2\rfloor$},\nonumber \\ &\phantom{{}==={}}\text{$\vec{t}\in\!\{[b_0,b_1,b_2]\,|\,\forall{j},b_j\!\!\in\!\mathbb{Z}_{>0}\text{ and }N\!\leq\!N_\mathrm{max}\}$},
\end{align}
where $\mathbb{Z}_{>0}$ is the set of all positive integers. We chose the quantity $\lfloor m_\mathrm{tot}/2\rfloor$ as the maximum number of \typeii{} nodes in our optimization since the approximations that make \cref{eqn:SKR-secret-key-rate} accurate starts to break down beyond this value. This is because we can no longer properly place the \typeii{} nodes such that they are evenly spaced beyond this value (see \supplmat{}).
The rest of the parameters are fixed and their values are shown in \cref{tab:contants}.
To ensure that the true optimum does not lie in regime where only \typeii{} nodes are permitted in the network, we independently optimized such a configuration with respect to the cost function and found that both the resulting secret key rate and cost are significantly worse than for the hybrid configuration (see \supplmat{}).
 
\section*{Data availability}
The generated data in this study can be found at \url{https://doi.org/10.4121/b9c7327e-97b2-4ea2-9b74-18c51f265027.v1}.

\section*{Code availability}
The code used for obtaining the presented numerical results is available at \url{https://github.com/bernwo/code-concatenated-quantum-repeater}.

\section*{Acknowledgments}
We thank Stephanie Wehner for helpful discussions. K.J.W., J.B., and G.A. acknowledge funding from the NWO Gravitation Program Quantum Software Consortium (Project QSC No.~024.003.037). G.P. and T.S. were supported by the Federal Ministry of Education and Research (BMBF, Project QR.X with Subproject No.~KIS6QK4001, Project QPIS No.~16KISQ032K) and the European Research Council (ERC Starting Grant ``QUREP''). M.F.M.R. was supported by the Austrian Science Fund (FWF) through projects No. P36009-N and No. P36010-N and Finanziert von der Europ{\"a}ischen Union - NextGenerationEU. A.S.S. acknowledge funding from the Danish Nation Research Foundation (Center of Excellence ``Hy-Q,'' Grant No.~DNRF139). F.R. and L.J. acknowledge support from the ARO(W911NF-23-1-0077), ARO MURI (W911NF-21-1-0325), AFOSR MURI (FA9550-19-1-0399, FA9550-21-1-0209), AFRL (FA8649-21-P-0781), NSF (OMA-1936118, ERC-1941583, OMA-2137642), NTT Research, Packard Foundation (2020-71479), and the Marshall and Arlene Bennett Family Research Program.

\section*{Competing Interests}
The authors declare no competing interests.

\section*{Author Contributions}
J.B. conceived the quantum-repeater-network architecture together with A.S.S. J.B. supervised the project. K.J.W. developed theory and code for performing the numerical optimization of repeater network with input from A.S.S., F.R., and L.J., while G.A. and M.F.M.R developed the code for finding the effective error suppression provided by specific tree branching vectors. All authors contributed to the discussion of the results and writing of the manuscript.

\section*{Additional information}
\textbf{Supplemental material} The online version contains supplemental material available at \supplmatlink{}.

\textbf{Correspondence} and requests for materials should be addressed to K.J.W., G.A., or J.B.

\end{document}


\title{Supplemental material: Resource efficient fault-tolerant one-way quantum repeater with code concatenation}
\maketitle

\section{Error/erasure correction protocol in a \typeii{} node}\label{sec:qec-protocol-typeii-node}
As explained in the main text, a message qubit in some arbitrary state is first encoded into a logical qubit on the 5-qubit~code level, which comprises 5 physical spin qubits.
Each of those 5 physical spin qubits is then encoded onto a tree cluster state which are sent along the repeater network in parallel across 5 sets of optical fibers.
The repeater network consists of both \typei{} and \typeii{} nodes. Once the 5 sets of trees arrive at a \typeii{} node, it performs quantum error correction at the 5-qubit~code level in a fault-tolerant manner via flag qubits.
However, if a tree happens to be completely unrecoverable, i.e., more than 50\% of the photons in the tree were lost in transmission, then quantum erasure correction, which does not utilize the flag qubits, is performed instead. The protocol is shown as a flowchart in \cref{fig:individual-repeater-protocol-flowchart} below.

\begin{figure}[!h]
    \centering
    \includegraphics[width=0.475\linewidth]{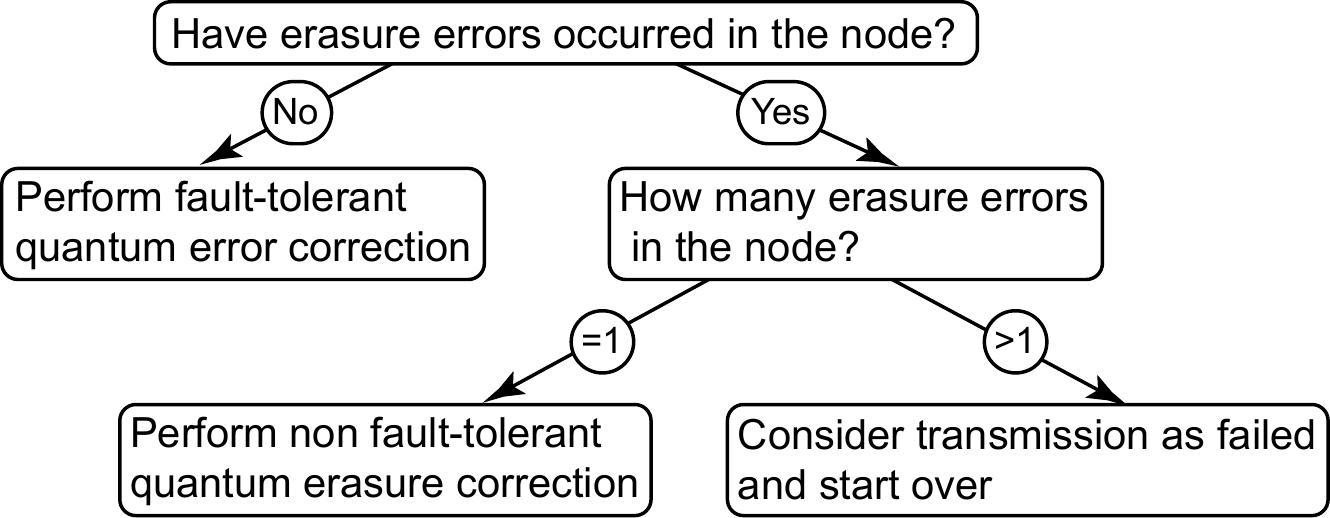}
    \caption{The control flow diagram of how a \typeii{} node decides whether to perform quantum error correction or quantum erasure correction.}
    \label{fig:individual-repeater-protocol-flowchart}
\end{figure}
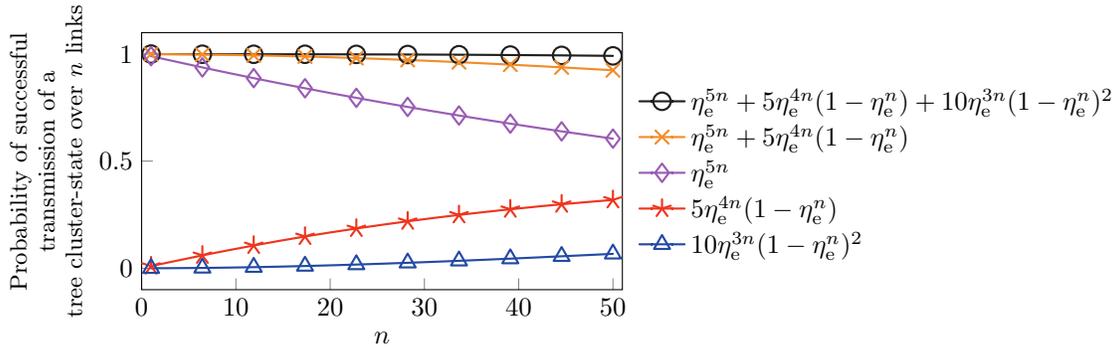
\begin{figure}[b!]
    \centering
\clearpage{}%
\begin{tikzpicture}
    \begin{axis}[
            xlabel={$n$},
            ylabel style={align=center,font=\small},
            ylabel=Probability of successful\\transmission of a\\tree cluster-state over $n$ links,
            xmin=0,
            xmax=51,
            ymin=-0.1,
            ymax=1.1,
            legend style={at={(2.05,0.825)},anchor=north east,draw=none,fill=none,font=\normalsize},
            legend cell align={left},
            grid=none,
        ]
        \addplot [mybla,mark=o,mark size=3.3pt, thick, line cap=round,line join=round,samples=10,smooth,domain=1:50] {(0.998)^(5*x)+5*(0.998)^(4*x)*(1-(0.998)^x)+10*(0.998)^(3*x)*(1-(0.998)^x)^2};
        \addplot [myora,mark=x,mark size=3.5pt, thick, line cap=round,line join=round,samples=10,smooth,domain=1:50] {(0.998)^(5*x)+5*(0.998)^(4*x)*(1-(0.998)^x)};
        \addplot [mypur,mark=diamond,mark size=3.5pt, thick, line cap=round,line join=round,samples=10,smooth,domain=1:50] {(0.998)^(5*x)};
        \addplot [myred,mark=star,mark size=3.5pt, thick, line cap=round,line join=round,samples=10,smooth,domain=1:50] {5*(0.998)^(4*x)*(1-(0.998)^x)};
        \addplot [myblu,mark=triangle,mark size=3.5pt, thick, line cap=round,line join=round,samples=10,smooth,domain=1:50] {10*(0.998)^(3*x)*(1-(0.998)^x)^2};
        \addlegendentry{$\etae^{5n}+5\etae^{4n}(1-\etae^n)+10\etae^{3n}(1-\etae^n)^2$}
        \addlegendentry{$\etae^{5n}+5\etae^{4n}(1-\etae^n)$}
        \addlegendentry{$\etae^{5n}$}
        \addlegendentry{$5\etae^{4n}(1-\etae^n)$}
        \addlegendentry{$10\etae^{3n}(1-\etae^n)^2$}
    \end{axis}
\end{tikzpicture}
\clearpage{}%

    \caption{
        The probabilities of successful transmission of a tree cluster-state over $n$ links as a function of $n$.
        The transmission probability of a tree cluster-state over 1 link is given by $\etae$ (see main text).
        The circle symbol denotes the total successful transmission probability assuming that both the 1- and 2-erasure errors are being corrected.
        The cross symbol denotes the total successful transmission probability assuming that only the 1-erasure errors are being corrected.
        The contributions to the total probability are shown by the diamond, star, and triangle symbols which denote the probability of no trees being lost, one tree being lost, and two trees being lost, respectively.
        We fixed the efficiency to $\etae=0.998$, which is the typical value found in our numerical optimization.
    }
    \label{fig:etae-vs-n}
\end{figure}
As shown in \cref{fig:individual-repeater-protocol-flowchart}, the \typeii{} node performs the non fault-tolerant quantum erasure correction when a 1-erasure error is heralded.
This happens with probability $5\etae^{4n}(1-\etae^n)$, where $\etae$ is the transmission probability of a tree from one node to the next and $n-1$ is the number of \typei{} nodes between consecutive \typeii{} nodes as explained in the main text.
However, for $k$-erasure errors with $k>1$, we abort and restart the transmission of the message qubit. This happens with probability $1-\etae^{5n}-5\etae^{4n}(1-\etae^n)$.
In theory, a \typeii{} node operating under the 5-qubit~code is also capable of correcting for 2-erasure errors as explained in the main text.
We did not consider this in our secret key rate analysis because the probability of two trees being unrecoverable at the same \typeii{} node is negligible compared to the probability for a 1-erasure error.
To show this, we first note that for the data points where the secret key rate of the hybrid repeater scheme is higher than that of the homogeneous repeater scheme (see main text) and $\epsilonr\in\{0.3\text{\textperthousand},0.5\text{\textperthousand},0.1\%,0.2\%\}$, the typical number of links is found to be in the range of $1\!\lesssim\!n\!\lesssim\!50$.
Plotting the individual contributing terms to the total successful transmission probability for this range in \cref{fig:etae-vs-n} reveals that the contributing term from the 2-erasure error correction is indeed negligible.
Note that this term is also negligible for when $\epsilonr=0.1\text{\textperthousand}$ where the corresponding number of links found is $50\!\lesssim\!n\!\lesssim\!150$ because the $\etae$ in this regime is much closer to unity compared to other higher re-encoding error probabilities.
By omitting the contribution from correcting 2-erasure errors, we are slightly underestimating our secret key rate.

\subsection{Fault-tolerant error correction protocol}\label{subsec:fault-tolerant-error-correction}
\begin{figure}[b]
    \centering
    \includegraphics[width=0.995\linewidth]{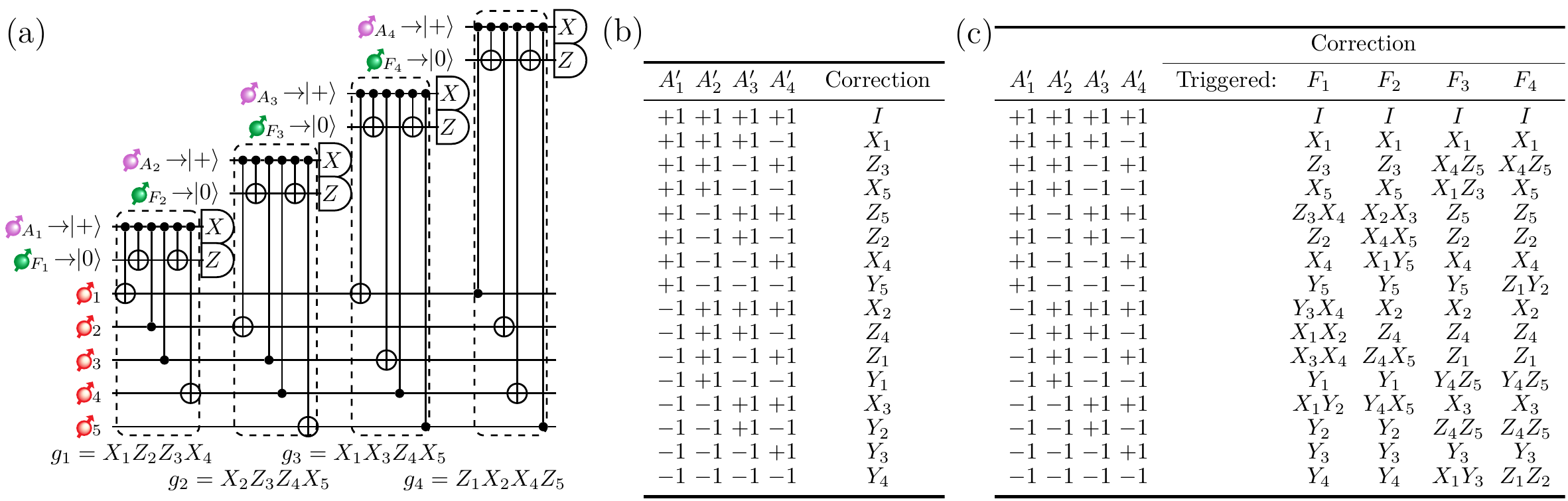}
    \caption{
        (a) The fault-tolerant 5-qubit~code quantum circuit with flag qubit.
        The same ancilla (flag) spin is re-initialized in state $\ket{+}$ ($\ket{0}$) before being sequentially measured at each stabilizer $g_k$ for $k\in\{1,2,3,4\}$.
        Sections of the circuit, i.e., sub-circuits, are boxed and labelled according to their respective stabilizers $g_k$ acting on the data qubits $\redspin_j$ with $j\in\{1,2,3,4,5\}$.
        (b) Correction look-up table of weight $w\leq1$ and (c) table of weight $w\leq2$.
    }
    \label{fig:five_qubit_code_flagged_qcircuit}
\end{figure}

Here, we explain the details of the fault-tolerant quantum error correction protocol from Ref.~\cite{Chao2018} as noted in both the main text and \cref{fig:individual-repeater-protocol-flowchart}.
The quantum circuit corresponding to the fault-tolerant 5-qubit~code error correction is shown in \figref{fig:five_qubit_code_flagged_qcircuit}{a} along with the sub-circuits labelled according to their respective stabilizers, which as a reminder are $X_1Z_2Z_3X_4$, $X_2Z_3Z_4X_5$, $X_1X_3Z_4Z_5$, and $Z_1X_2X_4Z_5$.
The protocol for the fault-tolerant 5-qubit~code error correction is shown explicitly in \cref{def:fault-tolerant-error-correction-protocol}. This protocol enables us to suppress weight $w=2$ errors which otherwise would have caused undetectable logical errors.
The protocol involves exiting from the fault-tolerant circuit under certain conditions, and changing into the non fault-tolerant circuit, i.e., the 5-qubit~code circuit without the flag qubit, to extract the syndrome from the ancilla qubits.
To differentiate the syndrome extracted from the fault-tolerant circuit from the non fault-tolerant circuit, we label the syndrome extracted from the non fault-tolerant circuit with the ``$\prime$'' sign, i.e., $A_k'$, which represents the value of the $k^\mathrm{th}$ measured ancilla qubit.

\begin{definition}\label{def:fault-tolerant-error-correction-protocol}
    The protocol for the fault-tolerant 5-qubit~code is as follows:

    Begin with the fault-tolerant circuit in \cref{fig:five_qubit_code_flagged_qcircuit} and proceed to performing the sub-circuits in order.
    Depending on the measurement outcome at each sub-circuit denoted by $g_k$ for $k\in\{1,2,3,4\}$, one would need to make the following decisions:
    \begin{enumerate}
        \itemsep-0.32em
        \item If $F_k=A_k=+1$, i.e., flag and ancilla are not triggered, then continue using the fault-tolerant circuit and extract $A_{k+1}$ and $A_{k+1}$. If all four flag qubits and four ancilla qubits were not triggered, we are finished with no corrections needed.
        \item If $F_k=+1$ and $A_k=-1$, i.e., flag is not triggered while ancilla is triggered, then switch to the non fault-tolerant circuit and measure $A_1', A_2',A_3',A_4'$. Finish by applying the weight $w\leq1$ corrections shown in \figref{fig:five_qubit_code_flagged_qcircuit}{b}.
        \item If $F_k=-1$ and $A_k=\pm1$, i.e., flag is triggered regardless of syndrome outcome, then switch to the non fault-tolerant circuit and measure $A_1', A_2', A_3', A_4'$. Finish by applying the weight $w\leq2$ corrections shown in \figref{fig:five_qubit_code_flagged_qcircuit}{c}.
    \end{enumerate}
\end{definition}%

\subsection{Effect of erasure correction on the 5-qubit~code level}
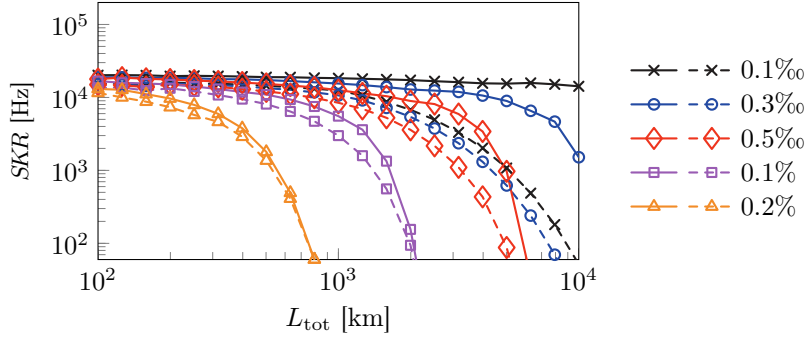
\begin{figure}[t]
    \centering
\clearpage{}%
\begin{tikzpicture}
    \begin{loglogaxis}[
            legend columns=2,
            xlabel={$L_\mathrm{tot}$ [km]},
            ylabel style={align=center,font=\small},
            ylabel={$\SKR$ [Hz]},
            xmin=10^2,
            xmax=10^4,
            ymin=6*10^1,
            ymax=2*10^5,
            legend style={at={(1.5,0.825)},anchor=north east,draw=none,fill=none,font=\normalsize},
            legend cell align={left},
            grid=none,
        ]
        \addplot [mybla,mark=x,mark size=2.5pt, thick, line cap=round,line join=round] table [x index=0, y expr=\thisrowno{1}, header=true, col sep=comma] {assets/csv/21_points_ea49e5c8/SKR_kappa0001/SKR_reencodingError0.0001_kappa0001.csv};
        \addplot [mybla,mark=x,mark size=2.5pt, thick,dashed, line cap=round,line join=round, mark options=solid] table [x index=0, y expr=\thisrowno{1}, header=true, col sep=comma] {assets/csv/21_points_a0f8a1b4/SKR_kappa0001_noErasure/SKR_reencodingError0.0001_kappa0001_noErasure.csv};
        \addplot [myblu,mark=o,mark size=2.0pt, thick, line cap=round,line join=round] table [x index=0, y expr=\thisrowno{1}, header=true, col sep=comma] {assets/csv/21_points_ea49e5c8/SKR_kappa0001/SKR_reencodingError0.0003_kappa0001.csv};
        \addplot [myblu,mark=o,mark size=2.0pt, thick,dashed, line cap=round,line join=round, mark options=solid] table [x index=0, y expr=\thisrowno{1}, header=true, col sep=comma] {assets/csv/21_points_a0f8a1b4/SKR_kappa0001_noErasure/SKR_reencodingError0.0003_kappa0001_noErasure.csv};
        \addplot [myred,mark=diamond, mark size=4pt, thick, line cap=round,line join=round] table [x index=0, y expr=\thisrowno{1}, header=true, col sep=comma] {assets/csv/21_points_ea49e5c8/SKR_kappa0001/SKR_reencodingError0.0005_kappa0001.csv};
        \addplot [myred,mark=diamond, mark size=4pt, thick,dashed, line cap=round,line join=round, mark options=solid] table [x index=0, y expr=\thisrowno{1}, header=true, col sep=comma] {assets/csv/21_points_a0f8a1b4/SKR_kappa0001_noErasure/SKR_reencodingError0.0005_kappa0001_noErasure.csv};
        \addplot [mypur,mark=square,mark size=1.7pt, thick, line cap=round,line join=round] table [x index=0, y expr=\thisrowno{1}, header=true, col sep=comma] {assets/csv/21_points_ea49e5c8/SKR_kappa0001/SKR_reencodingError0.001_kappa0001.csv};
        \addplot [mypur,mark=square,mark size=1.7pt, thick,dashed, line cap=round,line join=round, mark options=solid] table [x index=0, y expr=\thisrowno{1}, header=true, col sep=comma] {assets/csv/21_points_a0f8a1b4/SKR_kappa0001_noErasure/SKR_reencodingError0.001_kappa0001_noErasure.csv};
        \addplot [myora,mark=triangle,mark size=2.5pt, thick, line cap=round,line join=round] table [x index=0, y expr=\thisrowno{1}, header=true, col sep=comma] {assets/csv/21_points_ea49e5c8/SKR_kappa0001/SKR_reencodingError0.002_kappa0001.csv};
        \addplot [myora,mark=triangle,mark size=2.5pt, thick,dashed, line cap=round,line join=round, mark options=solid] table [x index=0, y expr=\thisrowno{1}, header=true, col sep=comma] {assets/csv/21_points_a0f8a1b4/SKR_kappa0001_noErasure/SKR_reencodingError0.002_kappa0001_noErasure.csv};
        \addlegendentry{}
        \addlegendentry{$0.1\text{\textperthousand}$}
        \addlegendentry{}
        \addlegendentry{$0.3\text{\textperthousand}$}
        \addlegendentry{}
        \addlegendentry{$0.5\text{\textperthousand}$}
        \addlegendentry{}
        \addlegendentry{$0.1\%$}
        \addlegendentry{}
        \addlegendentry{$0.2\%$}

    \end{loglogaxis}
\end{tikzpicture}
\clearpage{}%

    \caption{
        The secret key rate $\SKR$ of the hybrid repeater network for different values of re-encoding error rates $\epsilonr$ optimized with respect to the cost function shown in the main text. The solid and dashed lines represent the $\SKR$ values when the 5-qubit code is and is not leveraged for erasure error correction, respectively.
    }
    \label{fig:erasure-vs-noerasure-correction}
\end{figure}

In this section, we show that by leveraging the 5-qubit code to additionally correct for erasure errors, we obtain a substantial increase in the resulting optimized secret key rate. This increase is shown in \cref{fig:erasure-vs-noerasure-correction}.
We can also see that the higher re-encoding errors limit the potential for improved secret key rates. This is expected because for a 1-erasure error correction to succeed, it is imperative that the rest of the qubits are error free. Once the error rate is too high, we lose the advantage of 1-erasure error correction.

\subsection{Processing time in a \typeii{} node}\label{subsec:processing-time}
To quantify the secret key rate of the network, it is important to consider the processing time per repeater node. Since a \typeii{} node would always take longer than a \typei{} node for processing and become the processing bottleneck, we analyze only the processing time of a \typeii{} node.
Let $p_j$ represent the probability that either the ancilla or the flag qubit is detected as triggered in the readout of the $j^{\mathrm{th}}$ sub-circuit in \figref{fig:five_qubit_code_flagged_qcircuit}{a}, and $p_5$ represents the probability that no ancilla and flag qubits were triggered in the entirety of the fault-tolerant circuit. Note that $\sum_{j=1}^{5}p_j=1$ must be true.
Then, the act of performing fault-tolerant quantum error correction takes a time
\begin{equation}\label{eqn:T_f}
    T_\mathrm{f}=p_5\sum^{4}_{i=1}\tau_{\mathrm{f}_i}
    +
    \sum_{k=1}^{4}p_k\Big(\tau_\mathrm{nf}+\sum^{k}_{l=1}\tau_{\mathrm{f}_l}\Big),
\end{equation}
where $\tau_\mathrm{nf}=3\tau_\mathrm{ss}+13\tau_\mathrm{tele}+4\tau_\mathrm{meas}$ is the time required to perform the non fault-tolerant circuit while $\tau_{\mathrm{f}_k}$ is the time needed to perform the sub-circuits shown in \figref{fig:five_qubit_code_flagged_qcircuit}{a}, and they are given by

\begin{align}
    \tau_{\mathrm{f}_1} & =3\tau_{\mathrm{ss}}+3\tau_{\mathrm{tele}}+\tau_{\mathrm{meas}},\qquad
    \tau_{\mathrm{f}_2} =2\tau_{\mathrm{ss}}+4\tau_{\mathrm{tele}}+\tau_{\mathrm{meas}},\nonumber \\
    \tau_{\mathrm{f}_3} & =3\tau_{\mathrm{ss}}+3\tau_{\mathrm{tele}}+\tau_{\mathrm{meas}},\qquad
    \tau_{\mathrm{f}_4} =3\tau_{\mathrm{ss}}+3\tau_{\mathrm{tele}}+\tau_{\mathrm{meas}}.
\end{align}
It follows then that the total time to perform all subcircuits is $\sum_{l=1}^{4} \tau_{\mathrm{f}_{l}}=11\tau_{\mathrm{ss}}+13\tau_{\mathrm{tele}}+4\tau_{\mathrm{meas}}$.
Note that the maximum of $T_{\mathrm{f}}$ occurs when $p_4=1$. This corresponds to the situation in which no subcircuits were skipped in \figref{fig:five_qubit_code_flagged_qcircuit}{a}. If any \typeii{} node takes this long for processing, then the rest of the nodes in the network need to wait for just as long, thus it becomes the bottleneck of the network processing time. It is this time that we consider in the total processing time of repeater nodes in the network together with the tree generation time $\tau_{\mathrm{tree}}$, which is explained in the main text. The total processing time per node is thus given by
\begin{align}
    \tau_{\mathrm{tot}} & =  \tau_{\mathrm{tree}}+\tau_{\mathrm{nf}}+\sum_{l=1}^{4} \tau_{\mathrm{f}_l},\nonumber \\
                        & = \tau_{\mathrm{tree}}+14\tau_{\mathrm{ss}}+26\tau_\mathrm{tele}+8\tau_\mathrm{meas}.
\end{align}

\section{Maximized secret key rate}
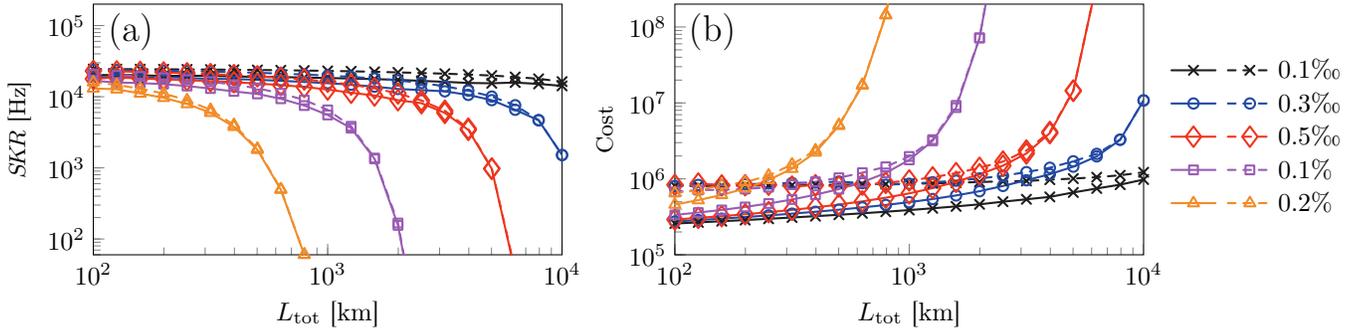
\begin{figure}[t]
    \centering
    \scalebox{0.975}{
\clearpage{}%
\begin{tikzpicture}
  \begin{loglogaxis}[
      title=(a),
      title style={at={(0.08,0.72)},font=\Large},
      legend columns=2,
      xlabel={$L_\mathrm{tot}$ [km]},
      ylabel style={align=center,font=\small},
      ylabel={$\SKR$ [Hz]},
      xmin=10^2,
      xmax=10^4,
      ymin=6*10^1,
      ymax=2*10^5,
      grid=none,
    ]
    \addplot [mybla,mark=x,mark size=2.5pt, thick, line cap=round,line join=round] table [x index=0, y expr=\thisrowno{1}, header=true, col sep=comma] {assets/csv/21_points_ea49e5c8/SKR_kappa0001/SKR_reencodingError0.0001_kappa0001.csv};
    \addplot [mybla,mark=x,mark size=2.5pt, thick,dashed, line cap=round,line join=round, mark options=solid] table [x index=0, y expr=\thisrowno{1}, header=true, col sep=comma] {assets/csv/21_points_dc4bd9b3/SKR_kappa0001_maxSKR/SKR_reencodingError0.0001_kappa0001_maxSKR.csv};
    \addplot [myblu,mark=o,mark size=2.0pt, thick, line cap=round,line join=round] table [x index=0, y expr=\thisrowno{1}, header=true, col sep=comma] {assets/csv/21_points_ea49e5c8/SKR_kappa0001/SKR_reencodingError0.0003_kappa0001.csv};
    \addplot [myblu,mark=o,mark size=2.0pt, thick,dashed, line cap=round,line join=round, mark options=solid] table [x index=0, y expr=\thisrowno{1}, header=true, col sep=comma] {assets/csv/21_points_dc4bd9b3/SKR_kappa0001_maxSKR/SKR_reencodingError0.0003_kappa0001_maxSKR.csv};
    \addplot [myred,mark=diamond, mark size=4pt, thick, line cap=round,line join=round] table [x index=0, y expr=\thisrowno{1}, header=true, col sep=comma] {assets/csv/21_points_ea49e5c8/SKR_kappa0001/SKR_reencodingError0.0005_kappa0001.csv};
    \addplot [myred,mark=diamond, mark size=4pt, thick,dashed, line cap=round,line join=round, mark options=solid] table [x index=0, y expr=\thisrowno{1}, header=true, col sep=comma] {assets/csv/21_points_dc4bd9b3/SKR_kappa0001_maxSKR/SKR_reencodingError0.0005_kappa0001_maxSKR.csv};
    \addplot [mypur,mark=square,mark size=1.7pt, thick, line cap=round,line join=round] table [x index=0, y expr=\thisrowno{1}, header=true, col sep=comma] {assets/csv/21_points_ea49e5c8/SKR_kappa0001/SKR_reencodingError0.001_kappa0001.csv};
    \addplot [mypur,mark=square,mark size=1.7pt, thick,dashed, line cap=round,line join=round, mark options=solid] table [x index=0, y expr=\thisrowno{1}, header=true, col sep=comma] {assets/csv/21_points_dc4bd9b3/SKR_kappa0001_maxSKR/SKR_reencodingError0.001_kappa0001_maxSKR.csv};
    \addplot [myora,mark=triangle,mark size=2.5pt, thick, line cap=round,line join=round] table [x index=0, y expr=\thisrowno{1}, header=true, col sep=comma] {assets/csv/21_points_ea49e5c8/SKR_kappa0001/SKR_reencodingError0.002_kappa0001.csv};
    \addplot [myora,mark=triangle,mark size=2.5pt, thick,dashed, line cap=round,line join=round, mark options=solid] table [x index=0, y expr=\thisrowno{1}, header=true, col sep=comma] {assets/csv/21_points_dc4bd9b3/SKR_kappa0001_maxSKR/SKR_reencodingError0.002_kappa0001_maxSKR.csv};
  \end{loglogaxis}
\end{tikzpicture}%
\begin{tikzpicture}
  \begin{loglogaxis}[
      title=(b),
      title style={at={(0.08,0.72)},font=\Large},
      legend columns=2,
      xlabel={$L_\mathrm{tot}$ [km]},
      ylabel style={align=center,font=\small},
      ylabel={Cost},
      xmin=10^2,
      xmax=10^4,
      ymin=10^5,
      ymax=2*10^8,
      legend style={at={(1.45,0.825)},anchor=north east,draw=none,fill=none,font=\normalsize},
      legend cell align={left},
      grid=none,
    ]
    \addplot [mybla,mark=x,mark size=2.5pt, thick, line cap=round,line join=round] table [x index=0, y expr=\thisrowno{1}, header=true, col sep=comma] {assets/csv/21_points_ea49e5c8/cost_kappa0001/cost_reencodingError0.0001_kappa0001.csv};
    \addplot [mybla,mark=x,mark size=2.5pt, thick,dashed, line cap=round,line join=round, mark options=solid] table [x index=0, y expr=\thisrowno{1}, header=true, col sep=comma] {assets/csv/21_points_dc4bd9b3/cost_kappa0001_maxSKR/cost_reencodingError0.0001_kappa0001_maxSKR.csv};
    \addplot [myblu,mark=o,mark size=2.0pt, thick, line cap=round,line join=round] table [x index=0, y expr=\thisrowno{1}, header=true, col sep=comma] {assets/csv/21_points_ea49e5c8/cost_kappa0001/cost_reencodingError0.0003_kappa0001.csv};
    \addplot [myblu,mark=o,mark size=2.0pt, thick,dashed, line cap=round,line join=round, mark options=solid] table [x index=0, y expr=\thisrowno{1}, header=true, col sep=comma] {assets/csv/21_points_dc4bd9b3/cost_kappa0001_maxSKR/cost_reencodingError0.0003_kappa0001_maxSKR.csv};
    \addplot [myred,mark=diamond, mark size=4pt, thick, line cap=round,line join=round] table [x index=0, y expr=\thisrowno{1}, header=true, col sep=comma] {assets/csv/21_points_ea49e5c8/cost_kappa0001/cost_reencodingError0.0005_kappa0001.csv};
    \addplot [myred,mark=diamond, mark size=4pt, thick,dashed, line cap=round,line join=round, mark options=solid] table [x index=0, y expr=\thisrowno{1}, header=true, col sep=comma] {assets/csv/21_points_dc4bd9b3/cost_kappa0001_maxSKR/cost_reencodingError0.0005_kappa0001_maxSKR.csv};
    \addplot [mypur,mark=square,mark size=1.7pt, thick, line cap=round,line join=round] table [x index=0, y expr=\thisrowno{1}, header=true, col sep=comma] {assets/csv/21_points_ea49e5c8/cost_kappa0001/cost_reencodingError0.001_kappa0001.csv};
    \addplot [mypur,mark=square,mark size=1.7pt, thick,dashed, line cap=round,line join=round, mark options=solid] table [x index=0, y expr=\thisrowno{1}, header=true, col sep=comma] {assets/csv/21_points_dc4bd9b3/cost_kappa0001_maxSKR/cost_reencodingError0.001_kappa0001_maxSKR.csv};
    \addplot [myora,mark=triangle,mark size=2.5pt, thick, line cap=round,line join=round] table [x index=0, y expr=\thisrowno{1}, header=true, col sep=comma] {assets/csv/21_points_ea49e5c8/cost_kappa0001/cost_reencodingError0.002_kappa0001.csv};
    \addplot [myora,mark=triangle,mark size=2.5pt, thick,dashed, line cap=round,line join=round, mark options=solid] table [x index=0, y expr=\thisrowno{1}, header=true, col sep=comma] {assets/csv/21_points_dc4bd9b3/cost_kappa0001_maxSKR/cost_reencodingError0.002_kappa0001_maxSKR.csv};
    \addlegendentry{}
    \addlegendentry{$0.1\text{\textperthousand}$}
    \addlegendentry{}
    \addlegendentry{$0.3\text{\textperthousand}$}
    \addlegendentry{}
    \addlegendentry{$0.5\text{\textperthousand}$}
    \addlegendentry{}
    \addlegendentry{$0.1\%$}
    \addlegendentry{}
    \addlegendentry{$0.2\%$}

  \end{loglogaxis}
\end{tikzpicture}
\clearpage{}%

    }
    \caption{
        (a) The secret key rate $\SKR$ of the hybrid repeater network with $\kappa=1$ for the configurations in which the secret key rate is optimized with respect to the cost function (solid lines) and purely maximized (dashed lines). (b) The corresponding cost.
    }
    \label{fig:optimise-vs-maxSKR}
\end{figure}

In this section, we discuss the difference of the secret key rate and cost between maximizing for the secret key rate (dashed lines) and optimizing the network configuration with respect to the cost function (solid lines) shown in \cref{fig:optimise-vs-maxSKR}.
In particular, we showed that in \figref{fig:optimise-vs-maxSKR}{a} that the discrepancy in the secret key rate is small between the two configurations, yet the cost is significantly higher for the configuration in which the secret key rate is maximized in \figref{fig:optimise-vs-maxSKR}{b}, especially at shorter distances.
This suggests that by merely optimizing the network with respect to the cost function, we can achieve near-optimum secret key rates at the benefit of significantly reduced resource requirement.
The secret key rate and cost converges however due to the increasing total distance, which limits the number of loss errors which can be corrected in the network.

\section{Having only \typeii{} nodes in the repeater network}
\begin{figure}[t]
    \centering
    \scalebox{0.975}{
\clearpage{}%
\begin{tikzpicture}
  \begin{loglogaxis}[
      title=(a),
      title style={at={(0.08,0.72)},font=\Large},
      legend columns=2,
      xlabel={$L_\mathrm{tot}$ [km]},
      ylabel style={align=center,font=\small},
      ylabel={$\SKR$ [Hz]},
      xmin=10^2,
      xmax=10^4,
      ymin=6*10^1,
      ymax=2*10^5,
      grid=none,
    ]
    \addplot [mybla,mark=x,mark size=2.5pt, thick, line cap=round,line join=round] table [x index=0, y expr=\thisrowno{1}, header=true, col sep=comma] {assets/csv/21_points_ea49e5c8/SKR_kappa0001/SKR_reencodingError0.0001_kappa0001.csv};
    \addplot [mybla,mark=x,mark size=2.5pt, thick,dashed, line cap=round,line join=round, mark options=solid] table [x index=0, y expr=\thisrowno{1}, header=true, col sep=comma] {assets/csv/21_points_9b98c5d0/SKR_kappa0001_full_typeII/SKR_reencodingError0.0001_kappa0001_full_typeII.csv};
    \addplot [myblu,mark=o,mark size=2.0pt, thick, line cap=round,line join=round] table [x index=0, y expr=\thisrowno{1}, header=true, col sep=comma] {assets/csv/21_points_ea49e5c8/SKR_kappa0001/SKR_reencodingError0.0003_kappa0001.csv};
    \addplot [myblu,mark=o,mark size=2.0pt, thick,dashed, line cap=round,line join=round, mark options=solid] table [x index=0, y expr=\thisrowno{1}, header=true, col sep=comma] {assets/csv/21_points_9b98c5d0/SKR_kappa0001_full_typeII/SKR_reencodingError0.0003_kappa0001_full_typeII.csv};
    \addplot [myred,mark=diamond, mark size=4pt, thick, line cap=round,line join=round] table [x index=0, y expr=\thisrowno{1}, header=true, col sep=comma] {assets/csv/21_points_ea49e5c8/SKR_kappa0001/SKR_reencodingError0.0005_kappa0001.csv};
    \addplot [myred,mark=diamond, mark size=4pt, thick,dashed, line cap=round,line join=round, mark options=solid] table [x index=0, y expr=\thisrowno{1}, header=true, col sep=comma] {assets/csv/21_points_9b98c5d0/SKR_kappa0001_full_typeII/SKR_reencodingError0.0005_kappa0001_full_typeII.csv};
    \addplot [mypur,mark=square,mark size=1.7pt, thick, line cap=round,line join=round] table [x index=0, y expr=\thisrowno{1}, header=true, col sep=comma] {assets/csv/21_points_ea49e5c8/SKR_kappa0001/SKR_reencodingError0.001_kappa0001.csv};
    \addplot [mypur,mark=square,mark size=1.7pt, thick,dashed, line cap=round,line join=round, mark options=solid] table [x index=0, y expr=\thisrowno{1}, header=true, col sep=comma] {assets/csv/21_points_9b98c5d0/SKR_kappa0001_full_typeII/SKR_reencodingError0.001_kappa0001_full_typeII.csv};
    \addplot [myora,mark=triangle,mark size=2.5pt, thick, line cap=round,line join=round] table [x index=0, y expr=\thisrowno{1}, header=true, col sep=comma] {assets/csv/21_points_ea49e5c8/SKR_kappa0001/SKR_reencodingError0.002_kappa0001.csv};
    \addplot [myora,mark=triangle,mark size=2.5pt, thick,dashed, line cap=round,line join=round, mark options=solid] table [x index=0, y expr=\thisrowno{1}, header=true, col sep=comma] {assets/csv/21_points_9b98c5d0/SKR_kappa0001_full_typeII/SKR_reencodingError0.002_kappa0001_full_typeII.csv};
  \end{loglogaxis}
\end{tikzpicture}%
\begin{tikzpicture}
  \begin{loglogaxis}[
      title=(b),
      title style={at={(0.08,0.72)},font=\Large},
      legend columns=2,
      xlabel={$L_\mathrm{tot}$ [km]},
      ylabel style={align=center,font=\small},
      ylabel={Cost},
      xmin=10^2,
      xmax=10^4,
      ymin=10^5,
      ymax=2*10^8,
      legend style={at={(1.45,0.825)},anchor=north east,draw=none,fill=none,font=\normalsize},
      legend cell align={left},
      grid=none,
    ]
    \addplot [mybla,mark=x,mark size=2.5pt, thick, line cap=round,line join=round] table [x index=0, y expr=\thisrowno{1}, header=true, col sep=comma] {assets/csv/21_points_ea49e5c8/cost_kappa0001/cost_reencodingError0.0001_kappa0001.csv};
    \addplot [mybla,mark=x,mark size=2.5pt, thick,dashed, line cap=round,line join=round, mark options=solid] table [x index=0, y expr=\thisrowno{1}, header=true, col sep=comma] {assets/csv/21_points_9b98c5d0/cost_kappa0001_full_typeII/cost_reencodingError0.0001_kappa0001_full_typeII.csv};
    \addplot [myblu,mark=o,mark size=2.0pt, thick, line cap=round,line join=round] table [x index=0, y expr=\thisrowno{1}, header=true, col sep=comma] {assets/csv/21_points_ea49e5c8/cost_kappa0001/cost_reencodingError0.0003_kappa0001.csv};
    \addplot [myblu,mark=o,mark size=2.0pt, thick,dashed, line cap=round,line join=round, mark options=solid] table [x index=0, y expr=\thisrowno{1}, header=true, col sep=comma] {assets/csv/21_points_9b98c5d0/cost_kappa0001_full_typeII/cost_reencodingError0.0003_kappa0001_full_typeII.csv};
    \addplot [myred,mark=diamond, mark size=4pt, thick, line cap=round,line join=round] table [x index=0, y expr=\thisrowno{1}, header=true, col sep=comma] {assets/csv/21_points_ea49e5c8/cost_kappa0001/cost_reencodingError0.0005_kappa0001.csv};
    \addplot [myred,mark=diamond, mark size=4pt, thick,dashed, line cap=round,line join=round, mark options=solid] table [x index=0, y expr=\thisrowno{1}, header=true, col sep=comma] {assets/csv/21_points_9b98c5d0/cost_kappa0001_full_typeII/cost_reencodingError0.0005_kappa0001_full_typeII.csv};
    \addplot [mypur,mark=square,mark size=1.7pt, thick, line cap=round,line join=round] table [x index=0, y expr=\thisrowno{1}, header=true, col sep=comma] {assets/csv/21_points_ea49e5c8/cost_kappa0001/cost_reencodingError0.001_kappa0001.csv};
    \addplot [mypur,mark=square,mark size=1.7pt, thick,dashed, line cap=round,line join=round, mark options=solid] table [x index=0, y expr=\thisrowno{1}, header=true, col sep=comma] {assets/csv/21_points_9b98c5d0/cost_kappa0001_full_typeII/cost_reencodingError0.001_kappa0001_full_typeII.csv};
    \addplot [myora,mark=triangle,mark size=2.5pt, thick, line cap=round,line join=round] table [x index=0, y expr=\thisrowno{1}, header=true, col sep=comma] {assets/csv/21_points_ea49e5c8/cost_kappa0001/cost_reencodingError0.002_kappa0001.csv};
    \addplot [myora,mark=triangle,mark size=2.5pt, thick,dashed, line cap=round,line join=round, mark options=solid] table [x index=0, y expr=\thisrowno{1}, header=true, col sep=comma] {assets/csv/21_points_9b98c5d0/cost_kappa0001_full_typeII/cost_reencodingError0.002_kappa0001_full_typeII.csv};
    \addlegendentry{}
    \addlegendentry{$0.1\text{\textperthousand}$}
    \addlegendentry{}
    \addlegendentry{$0.3\text{\textperthousand}$}
    \addlegendentry{}
    \addlegendentry{$0.5\text{\textperthousand}$}
    \addlegendentry{}
    \addlegendentry{$0.1\%$}
    \addlegendentry{}
    \addlegendentry{$0.2\%$}

  \end{loglogaxis}
\end{tikzpicture}
\clearpage{}%

    }
    \caption{
        (a) The secret key rate $\SKR$ of the hybrid repeater network with $\kappa=1$ for the configurations in which both types of nodes are allowed (solid lines) and only \typeii{} nodes are allowed (dashed lines). (b) The corresponding cost.
    }
    \label{fig:hybrid-vs-typeIIonly}
\end{figure}
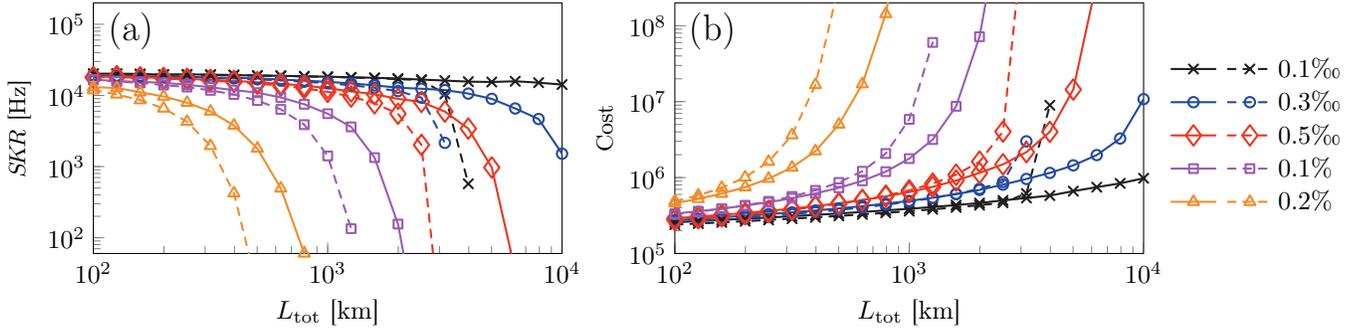

In this section, we show via \cref{fig:hybrid-vs-typeIIonly} that having only \typeii{} nodes in the network is not an optimal approach, despite optimizing for this configuration independently with respect to the cost function noted in the main text. Specifically, the resulting secret key rate in \figref{fig:hybrid-vs-typeIIonly}{a} is performing worse if we permit only \typeii{} nodes in the network.
The performance discrepancy is especially apparently at larger total distances, where loss errors are likely.
This inadvertently causes the cost in \figref{fig:hybrid-vs-typeIIonly}{b} to be higher.
This result implies that a code concatenated network in which only \typeii{} nodes is a configuration that is far from being ideal, especially at larger total distances.

\section{Approximating fidelity via recursion}
To calculate the resulting fidelity of our message qubit at the end node exactly, one can start with a density matrix of the error-free 5-qubit~code logical state, followed by the application of the stabilizer operations for each \typeii{} node in the network, effectively evolving the density matrix analytically. As noted in fig.~8 in the main text, however, there can be up to hundreds of \typeii{} nodes in the network, which makes the calculation a challenging task computationally. In this section, we introduce a novel method to approximate this fidelity up to great precision by exploiting the symmetry of the network.

Let $\rho_0$ be the 5-qubit density matrix that describes the 5-qubit~code logical state, free from any errors, and $\rho_\perp$ is a matrix, which attempts to describe the state that is orthogonal to the logical state.
The qubits in the network are subjected to errors in transmission, which is why a \typeii{} node performs (fault-tolerant) quantum error correction upon the reception of an incoming tree (as discussed in \cref{sec:qec-protocol-typeii-node}).
The effect of the errors introduced in the transmission and the quantum error correction performed by a \typeii{} node on the message qubit can be modelled as a quantum channel acting on the density matrix of the perfect logical state

\begin{equation}\label{eqn:quantum-channel-definition}
    \mathscr{C}(\rho_0)=\alpha_1\rho_0+\beta_1\rho_\perp,
\end{equation}
where $\alpha_1$ is the fidelity, $\beta_1=1-\alpha_1$, and the matrix $\rho_\perp$ is modelled as $\rho_\perp=(\mathscr{C}(\rho_0)-\alpha_1\rho_0)/\beta_1$.
As explained in the main text, we consider the nodes to be equidistant from each other regardless of their type. This means that we can apply the quantum channel in \cref{eqn:quantum-channel-definition} repeatedly, which reflects the fact that there are multiple \typeii{} nodes in the network with the same number of \typei{} nodes between them.
We model also the application of the quantum channel on the orthogonal state as

\begin{equation}\label{eqn:quantum-channel-on-orthogonal-state}
    \mathscr{C}(\rho_\perp)=\alpha'\rho_0+\beta'\rho_\perp'.
\end{equation}

By using \cref{eqn:quantum-channel-definition,eqn:quantum-channel-on-orthogonal-state}, we can obtain a \textit{recurrence relation} below.
As an approximation used in obtaining the recurrence relation, we assume that the correction on $\rho_\perp$ is highly efficient, i.e., $\alpha'\approx1$, and therefore we only keep terms that lead to non-zero fidelity while we discard terms with $\beta'$. Applying these rules to all stages of the protocol leads to

\begin{align}\label{eqn:recurrence-relation}
    \rho_1 = \mathscr{C}(\rho_0)                                                              & = \alpha_1\rho_0+\beta_1\rho_\perp,\nonumber    \\
    \rho_2 = \mathscr{C}(\rho_1) = \alpha_1\mathscr{C}(\rho_0)+\beta_1\mathscr{C}(\rho_\perp) & = \alpha_2\rho_0+\beta_2\rho_\perp,\nonumber    \\
                                                                                              & \vdots\nonumber                                 \\
    \rho_{\mII} = \mathscr{C}(\rho_{\mII-1})                                                  & = \alpha_{\mII}\rho_{0}+\beta_{\mII}\rho_\perp,
\end{align}
where $\alpha_{\mII}$ and $\beta_{\mII}$ are given by
\begin{equation}\label{eqn:alpha-beta-recurrence-relation}
    \alpha_{\mII}=\alpha_1\alpha_{\mII-1}+\beta_{\mII-1}\alpha',\quad\beta_{\mII}=\beta_1\alpha_{\mII-1}.
\end{equation}
From here, we obtain
\begin{equation}\label{eqn:combined-recurrence-relation}
    \alpha_{\mII}=\alpha_{\mII-1}\alpha_1+\alpha_{\mII-2}(1-\alpha_1)\alpha',\text{ with }\alpha_i=
    \begin{cases}
        \alpha_1, & \text{if } i=1 \\
        1,        & \text{if } i=0 \\
        0,        & \text{if } i<0
    \end{cases}.
\end{equation}
Finally, solving the recurrence relation in \cref{eqn:combined-recurrence-relation} yields
\begin{equation}\label{eqn:approximated-fidelity}
    \boxed{
        \mathscr{F}_{\mII,n}\equiv\alpha_{\mII}\equiv\frac{(\alpha_1+\zeta)^{\mII+1}-(\alpha_1-\zeta)^{\mII+1}}{2^{\mII+1}\zeta},\quad \mII\geq1,
    }
\end{equation}
where $\zeta=\sqrt{\alpha_1^2+4(1-\alpha_1)\alpha'}$ and $\mathscr{F}_{\mII,n}$ is the approximated fidelity for $\mII$ \typeii{} nodes and $n-1$ \typei{} nodes between consecutive \typeii{} nodes.
Note that the quantum channel $\mathscr{C}$ depends on $n$ and thus $\alpha_1$.
Hence, we label the approximated fidelity with the subscript $n$ in \cref{eqn:approximated-fidelity}.
The term $\alpha'$ can be expressed in terms of $\alpha_1$ and $\alpha_2$ by noting the expressions $\rho_1$ and $\rho_2$ in \cref{eqn:recurrence-relation}:

\begin{figure}[b]
    \centering
\clearpage{}%
\begin{tikzpicture}
    \begin{loglogaxis}[
            xlabel={$\epsilonr$},
            ylabel={$\epsiloneff$},
            xmin=0.0001,
            xmax=1,
            ymin=1*10^-4,
            ymax=1,
            legend style={at={(1.90,0.825)},anchor=north east,draw=none,fill=none,font=\normalsize},
            legend cell align={left},
            grid=none,
            ytick={10^-4,10^-3,10^-2,10^-1,1},
        ]
        \addplot [mybla, thick, line cap=round,line join=round] table [x index=0, y expr=\thisrowno{1}, header=true, col sep=comma] {assets/csv/exactError.csv};
        \addplot [myora, thick,dashed, line cap=round,line join=round] table [x index=0, y expr=\thisrowno{1}, header=true, col sep=comma] {assets/csv/recurError.csv};
        \addplot [myred, thick,dotted, line cap=round,line join=round] table [x index=0, y expr=\thisrowno{1}, header=true, col sep=comma] {assets/csv/pessiError.csv};
        \addplot [gray, line cap=round,line join=round] table [x index=0, y expr=\thisrowno{1}, header=true, col sep=comma] {assets/csv/elevenError.csv} node[pos=0.6,pin=325:{\color{gray}$12.61\%$}] {};
        \addlegendentry{Exact}
        \addlegendentry{Approximation via recursion}
        \addlegendentry{Na{\"i}ve approximation}
    \end{loglogaxis}
\end{tikzpicture}
\clearpage{}%

    \caption{
        The effective error rate $\epsiloneff$ as a function of the re-encoding error $\epsilonr$. The effective error rate was calculated exactly for $n=8$, $\mII=125$, assuming no erasure errors. The effective error rates are approximated using both the recursion method and na{\"i}ve method.
    }
    \label{fig:exact-vs-recur-effective-error-rates}
\end{figure}
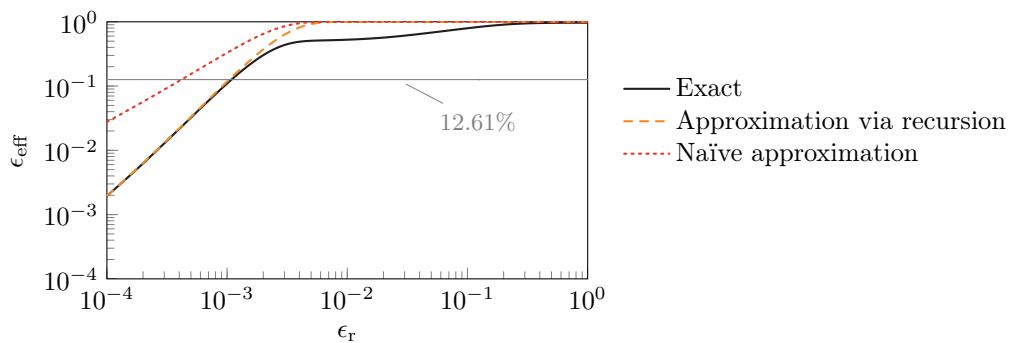

\begin{equation}\label{eqn:alpha-prime}
    \mathscr{C}(\rho_\perp)=\frac{\rho_2-\alpha_1\rho_1}{\beta_1}=\frac{\alpha_2+\alpha_1^2}{\beta_1}\rho_0+\cdots.
\end{equation}
By comparing \cref{eqn:alpha-prime,eqn:quantum-channel-on-orthogonal-state}, we find that $\alpha'=(\alpha_2+\alpha_1^2)/\beta_1$, and therefore we can express $\zeta$ in \cref{eqn:approximated-fidelity} as $\zeta=\sqrt{4\alpha_2-3\alpha_1^2}$.
We now have an efficient manner to obtain an approximated fidelity for $\mII>2$ by computing $\alpha_1$ and $\alpha_2$. Note that $\mathscr{F}_{\mII,n}$ as shown in \cref{eqn:approximated-fidelity} is exact for $\mII=1$ and $\mII=2$. Then, we can define the effective error as

\begin{equation}\label{eqn:effective-error-no-erasure}
    \epsiloneff=1-\mathscr{F}_{\mII,n}.
\end{equation}

Note that if we set $\alpha'=0$, we effectively arrive at the na{\"i}ve/pessimistic approximation, where the effective error simply becomes $1-\alpha^{\mII}_{1}$ shown in \cref{fig:exact-vs-recur-effective-error-rates}. In \cref{fig:exact-vs-recur-effective-error-rates}, we show that the approximation of using the recurrence relation (orange dashed line) yields remarkably accurate results compared to the exact effective error rate (solid black line) in the regime of interest, i.e., $\epsiloneff\lesssim12.61\%$ because $12.61\%$ is approximately the threshold QBER for generating secret keys in the six-state quantum key distribution protocol \cite{Scarani2009}, whereas the na{\"i}ve approximation (red dotted line) yields effective error rates about an order of magnitude higher than the exact result.

If erasure errors are also present, it is no longer feasible to only use the effective error in \cref{eqn:effective-error-no-erasure} because we the channels are no longer the same throughout the network. However, we can simply modify the expression to include correcting for $i$ 1-erasure errors in the network
\begin{equation}\label{eqn:effective-error-with-erasure}
    \epsiloneff(\mII,i)=1-(\mathscr{F}_{\mII-i,n})(1-\epsilon_{\mathrm{loss},n})^i.
\end{equation}
where $\epsilon_{\mathrm{loss},n}$ is defined as one minus the fidelity of the logical qubit at the 5-qubit~code level at a \typeii{} node after a 1-erasure correction with the condition that the logical qubit was error-free in the previous \typeii{} node with $n$ links between them. Note that it is implied in \cref{eqn:effective-error-with-erasure} that our message qubit cannot be recovered after 1 failed attempt at erasure correction at the 5-qubit~code level. This is a reasonable assumption since erasure correction is vulnerable to additional errors on the intact-qubits.

\subsection{Uneven distance between \typeii{} nodes}\label{subsec:uneven-distance-nodes}
It is not always possible to place \typeii{} nodes in the network such that they are evenly spaced between each other for a given total distance $L_\mathrm{tot}$. In this subsection, we explain the specific strategy we employed in the main text for placing the repeater nodes as evenly as possible.

\begin{figure}[b]
    \centering
    \includegraphics[width=0.65\linewidth]{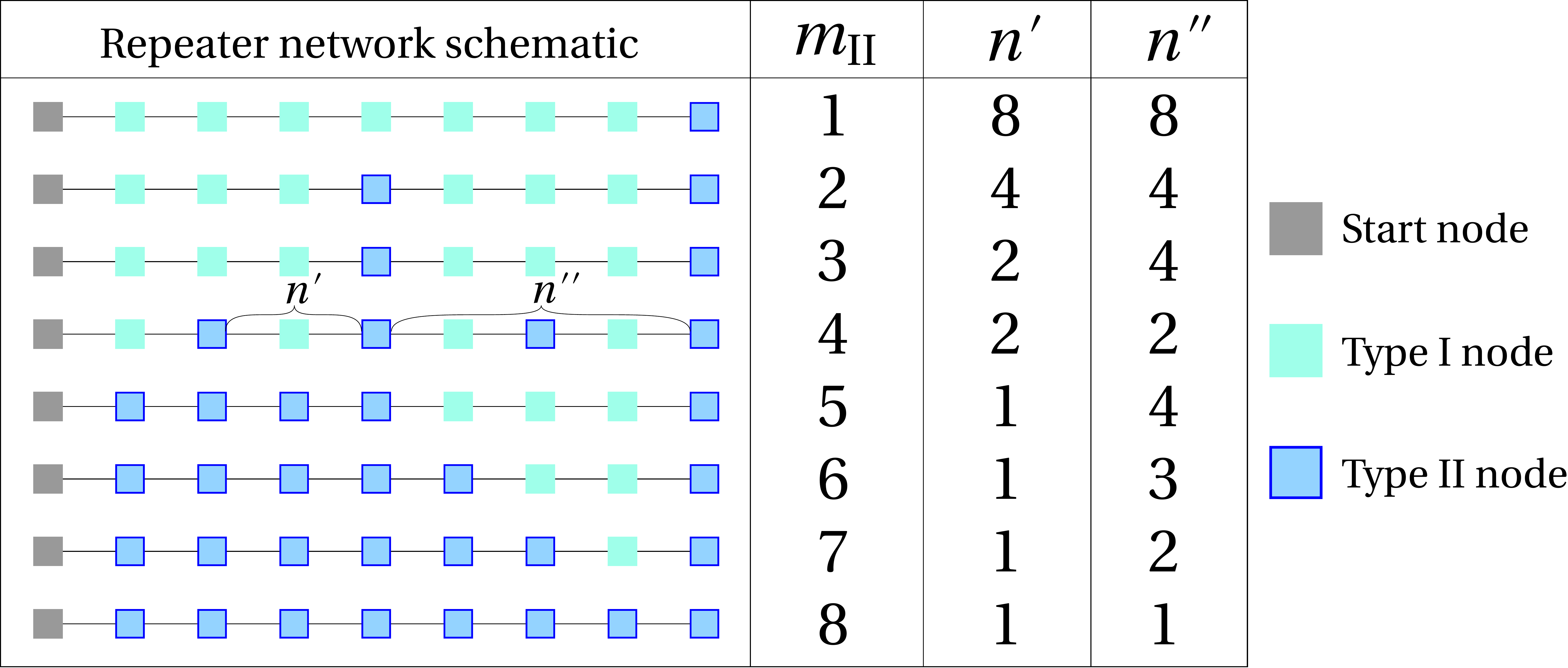}
    \caption{
        Repeater network layout in the presence of \typei{} and \typeii{} nodes for varying number of \typeii{} nodes $\mII$ and fixed number of total nodes $m_\mathrm{tot}=8$.
        Not all values of $\mII$ can divide $m_\mathrm{tot}$, hence there will be sections in which the number of links between consecutive \typeii{} nodes are different, i.e., $n'=\lfloor m_\mathrm{tot}/\mII\rfloor$ versus $n''=m_\mathrm{tot}-n'(\mII-1)$.
    }
    \label{fig:uneven-distance-typeii-nodes}
\end{figure}

For some given total distance $L_\mathrm{tot}$ we can calculate the total number of nodes in the network (excluding the start node) as
\begin{equation}\label{eqn:n-tot}
    m_\mathrm{tot}=\frac{L_\mathrm{tot}}{L_0},
\end{equation}
where $L_0$ is the distance between consecutive nodes.
Note that for convenience, we choose the values of $L_0$ such that $m_\mathrm{tot}\in\mathbb{Z}_{>0}$.
Given no erasure errors in the network, the effective error rate of the message qubit received at the end node is
\begin{equation}\label{eqn:effective-error-rate-zero-erasure-errors}
    \epsiloneff(\mII,0)=1-(\mathscr{F}_{\mII-1,n'})(\mathscr{F}_{1,n''}),
\end{equation}
where $n'=\lfloor m_\mathrm{tot}/\mII\rfloor$ and $n''=m_\mathrm{tot}-n'(\mII-1)$ are the number of links between consecutive \typeii{} nodes as shown in \cref{fig:uneven-distance-typeii-nodes}. Note that for $\mII>\lfloor m_{\mathrm{tot}}/2\rfloor$, the repeater nodes can no longer be placed reasonably evenly, thus we constrained our numerical minimization of the cost function to $\mII\leq\lfloor m_{\mathrm{tot}}/2\rfloor$ as noted in Methods section the main text.

\subsection{Accounting for erasure errors in the network}
Here, we modify \cref{eqn:effective-error-rate-zero-erasure-errors} to account for erasure error correction:
\begin{equation}
    \epsiloneff(\mII,i)\approx\begin{cases}
        1-(1-\epsilon_{\mathrm{loss},n})^{i}(\mathscr{F}_{\mII-i,m_\mathrm{tot}/\mII})        & ,\text{if }(m_\mathrm{tot}\!\!\!\!\!\mod \mII=0)                     \\
        1-(1-\epsilon_{\mathrm{loss},n'})^{i}(\mathscr{F}_{\mII-1-i,n'})(\mathscr{F}_{1,n''}) & ,\text{if }(m_\mathrm{tot}\!\!\!\!\!\mod \mII>0)\wedge(\mII-1\geq i) \\
        1-(1-\epsilon_{\mathrm{loss},n'})^{i-1}(1-\epsilon_{\mathrm{loss},n''})               & ,\text{if }(m_\mathrm{tot}\!\!\!\!\!\mod \mII>0)\wedge(\mII=i)\end{cases},
\end{equation}
This is the expression which is used in our numerical optimization of the secret key rate.

\section{Re-encoding error simulation}

In this section we determine the tree-level re-encoding error probability $\epsilonr$ and find that it relates to the single-qubit error probability $\epsilon_0$ according to
\begin{equation}
    \epsilonr \approx 3 \epsilon_0.
\end{equation}

First we introduce two different (yet equivalent) definitions of the single-qubit depolarizing channel
\begin{equation}
    \mathcal D_p (\rho) = p \rho + \frac{1 - p}{2} I = \mathcal D'_\epsilon(\rho) = (1 - \epsilon) \rho + \frac{\epsilon}{3} (X\rho X + Y \rho Y + Z \rho Z),
\end{equation}
where $X$, $Y$ and $Z$ are the Pauli operators, $I$ is the identity operator and
\begin{equation}
    \epsilon = \frac{3}{4} (1 - p).
\end{equation}
Here, we include both definitions because the first simplifies our treatment here while the second is the one used in the main text.
We model the re-encoding procedure as follows.
\begin{enumerate}
    \item
          The initial state $\rhoi$ is held by a memory qubit as result of heralded storage of the tree-encoded qubit transmitted from the previous node to the current node.
          A second memory qubit is the root of a fresh tree state (that is otherwise photonic) that is transmitted to the next node.

    \item
          The two memory qubits each undergo single-qubit depolarizing channel $\mathcal D_{p_0}$.

    \item
          A Bell-state measurement is performed on the two memory qubits to teleport the state $\rhoi$ into the fresh tree-level logical qubit.
          Single-qubit Paulis are applied to individual photons to realize the logical Pauli operation required by the teleportation protocol as specified by the outcome of the Bell-state measurement.

    \item
          Each of the photonic qubits is subject to both a single-qubit depolarizing channel $\mathcal D_{p_0}$ and a pure loss channel with loss probability $1 - \eta$.

    \item
          Heralded storage is attempted to put the transmitted tree-level logical qubit into a memory qubit again.
          Conditioned on success, the state held by the memory qubit is denoted $\rhof$.
\end{enumerate}
Here, $p_0$ is defined in terms of the single-qubit error probability according to $\epsilon_0 = \tfrac 3 4 (1 - p_0)$.
We will now set out to determine $\rhof$ as a function of $\rhoi$, $\epsilon_0$ and $\eta$.
\\

As each of the Bell states has the property $(I \otimes P) \ket{\phi} = \pm (P \otimes I) \ket{\phi}$ for all Pauli operators $P$, we can ``move'' the depolarizing noise occurring at the second memory qubit through the measurement operators of the Bell-state measurement to the first memory qubit.
Then, Step 3 is just the perfect teleportation of the state
\begin{equation}
    \mathcal D_{p_0} \circ \mathcal D_{p_0} (\rhoi)
\end{equation}
onto the tree-level logical qubit held by the photons.
Here, $\circ$ indicates composition of quantum channels.
Then, we define the quantum channel $\mathcal T_{p_0, \eta}(\rho)$ as the quantum channel implemented on the tree-level logical qubit by Steps 4 and 5.
The final state is therefore
\begin{equation}
    \rhof = \mathcal T_{p_0, \eta} \circ \mathcal D_{p_0} \circ \mathcal D_{p_0} (\rhoi).
\end{equation}

In order to determine $\mathcal T_{p_0, \eta}(\rho)$ we turn to numerical methods.
We use the quantum-network simulator NetSquid \cite{coopmans2021} to estimate the Choi state \cite{choi1975} of $\mathcal T_{p_0, \eta}(\rho)$.
We do so as follows.
First, we prepare the Bell state $\ket {\Phi^+} = \tfrac 1 {\sqrt 2} (\ket {00} + \ket{11})$.
Next, we perfectly teleport one of these two qubits onto a tree-level logical qubit (as done in Step 3).
Then, we subject each of the qubits in the tree to a pure loss channel with loss probability $1 - \eta$, and then to a depolarizing channel $\mathcal D_{p_0}$.
Finally, we simulate the heralded-storage procedure.
Conditioned on heralded storage being successful, the two-qubit state is now the Choi state
\begin{equation}
    \ket{\text{Choi}} = (\text{Id} \otimes \mathcal T_{p_0, \eta}) (\ket{\Phi^+}\bra{\Phi^+}).
\end{equation}
Here, Id is the identity quantum channel on a single qubit.
This state fully characterizes the quantum channel.
\\

Our simulations are probabilistic as each time a pure loss channel is applied a random number is sampled to determine if the photon is lost, and each time a depolarizing channel is applied, a random number is sampled to determine if there is an error (and if so, whether it is an X, Y or Z error).
For a range of values of $p_0$, $\eta$ and the branching vector of the tree code we have performed this procedure one million times to obtain reliable statistics.
The simulation code used to obtain the data can be found in \url{https://github.com/bernwo/code-concatenated-quantum-repeater}.
Both the raw and processed data can be found in \url{https://doi.org/10.4121/b9c7327e-97b2-4ea2-9b74-18c51f265027.v1}.
\\

We investigated this for all the parameter values $\vec{t}\in\{[4,13,4],[5,11,4],[4,14,4],[4,12,5]\}$ found via the numerical optimization of the secret key rate, and find that
\begin{equation}
    \ket{\text{Choi}} \approx (\text{Id} \otimes \mathcal D_{p_0}) (\ket{\Phi^+}\bra{\Phi^+})
\end{equation}
and thus
\begin{equation}
    \mathcal T_{p_0, \eta} \approx \mathcal D_{p_0}.
\end{equation}
This allows us to conclude
\begin{equation}
    \rhof \approx \mathcal D_{p_0} \circ \mathcal D_{p_0} \circ \mathcal D_{p_0} (\rhoi).
\end{equation}
Because $\mathcal D_{p_1} \circ \mathcal D_{p_2} = \mathcal D_{p_1 p_2}$ we then have
\begin{equation}
    \rhof \approx \mathcal D_{p_0^3} (\rhoi) \equiv \mathcal D'_{\epsilonr} (\rhoi).
\end{equation}
Here, the last part of the equation defines the re-encoding error probability $\epsilonr$.
Simply equating both sides then reveals
\begin{equation}
    \epsilonr = \frac 3 4 (1 - p_0^3) = \frac 3 4 \left(1 - \left( \frac 4 3 (1 - \epsilon_0) \right)^3 \right) = 3 \epsilon_0 + \mathcal O(\epsilon_0^2).
\end{equation}
We thus finally conclude that, for a small error probability $\epsilon_0 \ll 1$,
\begin{equation}
    \epsilonr \approx 3 \epsilon_0.
\end{equation}

\section{Comparison with other quantum repeater schemes}\label{sec:comparison}

There exists numerous proposals for quantum repeater architectures, which all seek to lower the hardware requirements for high-rate, long-distance quantum communication~\cite{Yin2017,Bhaskar2020,Pompili2021,Munro2015,Muralidharan2016,Briegel1998,Duan2001,Sangouard2011,Munro2012,Muralidharan2014,Ewert2016,Borregaard2020,Rozpedek2021,Fukui2021,Schmidt2022,Rozpedek2023, schmidt2023error}. Our protocol can be characterized as a one-way quantum repeater. This type of repeaters are particularly promising for hardware with an efficient spin-photon interface but limited memory performance both in terms of storage time and multiplexing capability.

The work of Ref.~\cite{Borregaard2020} showed that the use of photonic tree-cluster states in conjunction with efficient spin-photon interfaces could significantly reduce the qubit resources per repeater node compared with previous proposals. We refer the reader to the supplemental material of Ref.~\cite{Borregaard2020} for a detailed discussion of this aspect where comparison with both quantum emitter-based architectures~\cite{Muralidharan2014,Glaudell2016} and linear optics based approaches~\cite{Azuma2015,Ewert2017,Lee2019} has been performed.

We have summarized the comparison of our proposed quantum-repeater scheme in \cref{tab:comparison} with both the original proposal in Ref.~\cite{Borregaard2020}, the GKP-based protocols of Refs.~\cite{Fukui2021, Rozpedek2021,Rozpedek2023} and a recent photonic cluster state repeater proposal~\cite{Niu2022}. From the table, it is seen that our work significantly improves on previous work by allowing for a fault-tolerant repeater with minimized qubit resources per repeater node while representing a fundamentally different route than continuous variable GKP approaches.

\begin{table}[b!]
    \centering
    \begin{tabularx}{\linewidth}{l@{\hspace{1.0em}}s@{\hspace{1.0em}}j} \toprule[1.3pt]
        Schemes         & Characteristics & Performance       \\ \midrule \\ [-2.0ex]
        Current work    & \currMakecell   & \currPerformance  \\ \midrule
        \citet{Niu2022} & \niuMakecell    & \ciscoPerformance \\ \midrule
        \citeBorr       & \borrMakecell   & \borrPerformance  \\ \midrule
        \citeFukui      & \fukuiMakecell  & \fukuiPerformance \\
    \end{tabularx}
\end{table}%
\begin{table}[t!]
    \centering
    \begin{tabularx}{\linewidth}{l@{\hspace{1.0em}}s@{\hspace{1.0em}}j} \midrule
        \citeGKP    & \rozpMakecell    & \gkpPerformance    \\ \midrule
        \citeGKPnew & \rozpNewMakecell & \gkpNewPerformance \\\bottomrule[1.3pt]
    \end{tabularx}
    \caption{
        Comparison with quantum repeater schemes considered in Refs.~\cite{Niu2022,Borregaard2020,Rozpedek2021,Rozpedek2023}.
        As mentioned in the main text, $\eta_\mathrm{d}$ is the overall repeater efficiency, $L_0$ is the inter-repeater distance, $L_\mathrm{att}$ is the attenuation distance, and $L_\mathrm{tot}$ is the total distance.
    }
    \label{tab:comparison}
\end{table}%

The key drawback of Ref.~\cite{Borregaard2020} is that the lack of fault-tolerance which meant that very low error rates are needed for intercontinental distances. In this work, we have shown how this can be addressed efficiently through code concatenation and flag-based quantum error correction. Importantly, we demonstrate a fully fault-tolerant architecture with only a modest increase in resources compared to the protocol of Ref.~\cite{Borregaard2020}.

Recently, there has been a number of proposals that also exploit code-concatenation for fault-tolerant quantum repeater architectures~\cite{Rozpedek2021,Rozpedek2023, Fukui2021,Schmidt2022}.
These proposals focus on the combination of a GKP code with a small discrete-variable (DV) quantum error correcting code. Moreover, if one has access to almost lossless hardware, then in some configurations bare GKP qubits can be sufficient~\cite{Fukui2021} and in the limit of very high squeezing further performance benefits could be achieved by considering GKP qudits~\cite{schmidt2023error}.
These works represent a fundamentally different approach than the one we consider here.
In particular, the experimental challenges are very different for such approaches compared to ours. The generation of an optical GKP state remains a daunting experimental challenge, which is very different from  the arguably also challenging task of generating the large photonic cluster states that our protocol requires. We do note, however, that recent experimental achievements of the generation of multiple-photon cluster states have already demonstrated some of the key functionality required for our proposal~\cite{Thomas2022,Cogan2023}. We also note the recent experimental realization of optical GKP states~\cite{Konno2023} that provides the initial push towards the feasibility of GKP-based protocols.
In general, however, most of the systems currently being explored for quantum information processing are based on qubits. The DV approach that we pursue in this work, is thus applicable to a broader range of the experimental systems.